\documentclass{emulateapj}

\newcommand\uclchem{{\sc uclchem}~}

\shorttitle{The chemistry of phosphorus under energetic phenomena}
\shortauthors{Jim\'enez-Serra et al.}

\begin{document}

\title{The chemistry of phosphorus-bearing molecules under energetic phenomena}

\author{Izaskun Jim\'{e}nez-Serra\altaffilmark{1}, Serena Viti\altaffilmark{2}, David Qu\'enard\altaffilmark{1} and Jonathan Holdship\altaffilmark{2}}

\altaffiltext{1}{School of Physics \& Astronomy, Queen Mary University of London,
Mile End Road, E1 4NS, London (UK); i.jimenez-serra@qmul.ac.uk}
\altaffiltext{2}{Department of Physics \& Astronomy, University College London, Gower Street
London, WC1E 6BT London (UK)}

\begin{abstract}

For decades, the detection of phosphorus-bearing molecules in the interstellar medium was restricted to high-mass star-forming regions (as e.g. SgrB2 and Orion KL) and the circumstellar envelopes of evolved stars. However, recent higher-sensitivity observations have revealed that molecules such as PN and PO are present not only toward cold massive cores and low-mass star-forming regions with PO/PN ratios $\geq$1, but also toward the Giant Molecular Clouds in the Galactic Center known to be exposed to highly energetic phenomena such as intense UV radiation fields, shock waves and cosmic rays. In this paper, we carry out a comprehensive study of the chemistry of phosphorus-bearing molecules across different astrophysical environments which cover a range of physical conditions (cold molecular dark clouds, warm clouds, hot cores/hot corinos) and are exposed to different physical processes and energetic phenomena (proto-stellar heating, shock waves, intense UV radiation and cosmic rays). We show how the measured PO/PN ratio (either $\geq$1 as in e.g. hot molecular cores, or $\leq$1 as in UV strongly illuminated environments) can provide constraints on the physical conditions and energetic processing of the source. We propose that the reaction P + OH $\rightarrow$ PO + H, not included in previous works, could be an efficient gas-phase PO formation route in shocks. Our modelling provides a template with which to study the detectability of P-bearing species not only in regions in our own Galaxy but also in extragalactic sources.

\end{abstract}


\section{Introduction}

Phosphorus is considered as one of the main biogenic elements due to its key role in biochemical processes in living organisms \citep[][]{macia05}. Phosphorus is believed to be synthesised in massive stars and subsequently ejected into the interstellar medium (ISM) by supernovae explosions \citep{koo13}. Although its cosmic abundance is relatively high \citep[fractional abundance of $\sim$3$\times$10$^{-7}$ with respect to H;][]{asplund09}, it is difficult to detect in the ISM since it is likely to be heavily depleted onto dust grains \citep[see][]{turner90}.

Indeed, phosphorus-bearing species (hereafter P-bearing) have mainly been detected toward diffuse clouds \citep[in the form of P$^+$;][]{jura78} and circumstellar envelopes around evolved stars \citep[in the form of HCP, CP, CCP, NCCP, PO, PN and PH$_3$;][]{agundez07,agundez08,agundez14a,agundez14b,tenenbaum07,tenenbaum08,halfen08,milam08,debeck13}. More recently, higher-sensitivity observations have revealed the presence of P-bearing molecules such as PN and PO in regions with active low-mass and high-mass star formation \citep[][]{yamaguchi11,caux11,rivilla16,lefloch16}. PN has also been reported toward massive starless cores \citep[][]{fontani16}, which suggests that this molecule could form in relatively quiescent and cold gas \citep{mininni18}. In all cases, PO seems to be slightly more abundant than PN \citep[by factors 2-3; see e.g.][]{rivilla16}\footnote{Note that these PO/PN ratios were estimated assuming LTE conditions. These ratios may vary once collisional rate coefficients become available for PO with He and/or H$_2$ \citep[see][]{lique18}.}, and the PO/PN ratio can potentially be used as a length indicator of the pre-stellar collapse phase \citep[see][]{aota12,lefloch16}. The analysis of the P-bearing content of the Giant Molecular Clouds (GMCs) in the Galactic Center shows that PN and PO can also be detected in molecular clouds subject to extreme conditions such as those found in galactic nuclei \citep{rivilla18}. 

The lack of many detections of P-bearing molecules has indeed hampered the study of the chemistry of these species in the ISM. Initial theoretical works focused on the chemistry of cold and warm molecular clouds \citep{millar91} and massive hot molecular cores \citep{charnley94}. With the detection of P-bearing molecules in the circumstellar envelopes of evolved stars, some more modelling was carried out by \citet{mackay01} and \citet{agundez07}, exploring the non-equilibrium chemistry of phosphorus in the outer envelope of these objects. The recent detections of PN and PO in star-forming regions has triggered a new interest for the chemistry of phosphorus in the ISM. While \citet{yamaguchi11} and \citet{lefloch16} have proposed that P-bearing species are produced in shocked gas after the sputtering of dust grains, \citet{rivilla16} suggest that P-bearing molecules are formed during the cold collapse phase and subsequently frozen onto dust grains \citep{caux11,fontani16,rivilla16}. The detection of PN and PO in the Galactic Center poses even more unknowns to the chemistry of phosphorus \citep{rivilla18}. These authors have indeed found a positive trend between the abundance of PN and the optically-thin $^{29}$SiO isotopologue, which suggests that PN is generated in shocks. However, it is unclear whether P-bearing molecules can survive additional energetic processing due to the intense UV radiation and/or enhanced cosmic rays, known to be present in this region of the Galaxy.

In this paper, we present a comprehensive study of the chemistry of phosphorus in the ISM covering a wide range of physical conditions (from molecular dark clouds to cold/warm/hot cores) and of physical processes (protostellar heating, shock waves, UV radiation and cosmic rays). Our goal is to establish the most likely conditions under which P-bearing molecules form, and to what extent. Special focus is put on the analysis of the PO/PN abundance ratio to determine why this ratio is found to be  constant (within a factor of 2-3) across different sources. In Section 2, we describe the details of our chemical modelling, while in Section 3 we present the results. In Section 4, we discuss the implications of our modelling for Galactic and extragalactic studies, and we summarised our conclusions in Section 5. 
   
\section{Chemical modelling of P-bearing molecules}
\label{network}

To model the chemistry of P-bearing species, we have used the gas-grain chemical code \uclchem \citep{holdship17}\footnote{https://uclchem.github.io/}, which has been recently updated with a new treatment for the grain surface chemistry \citep{quenard18}. The code includes thermal and non-thermal desorption processes as described in \citet{viti04} and \citet{roberts07} respectively, and considers the grain surface processes of radical diffusion, chemical reactive desorption and reaction-diffusion competition as described in \citet{quenard18}. 

The chemical network is based on the one developed by \citet{quenard18}, which has been expanded to include P-bearing species. The gas-phase reactions were taken from the UMIST database  \citep{mcelroy13}, which includes the original network for phosphorus of \citet{millar91}. This network is based on the early experimental work of \citet{smith89} and \citet{thorne83,thorne84}, and only $\sim$30\% of the reactions have rate coefficients measured in the laboratory. We have also added the chemical reactions involving PH$_3$ from the chemical network of \citet[][]{charnley94}, who took the rate coefficients of the neutral-neutral reactions H + PH$_n$ ($n$=1-3) from \citet{kaye83} following the experimental data of \citet{lee76}. Other ion-neutral reactions involving PH$_n$ products and species such as PS or HPS were also included in our network following \citet{anicich93} and according to the experiments of \citet{smith89}.  
The two gas-phase reactions N+CP$\rightarrow$PN+C and P+CN$\rightarrow$PN+C proposed by \citet[][]{agundez07}, with reaction rates of 3$\times$10$^{-10}$$\,$cm$^3$s$^{-1}$, were also included in our network. Note that these reaction rates are highly uncertain since no experimental values are available. However, several models were run without these reactions and they yielded similar results.

For the grain surface chemistry, we have considered the hydrogenation reactions of P-bearing species into their non-saturated and saturated forms, as well as their associated diffusion and chemical reaction desorption reactions. Binding energies were taken from the KIDA database \citep{wakelam17}\footnote{http://kida.obs.u-bordeaux1.fr/}. In total, the network contains 3695 reactions involving 401 species, 265 out of which are gas phase species while the remaining 136 are grain surface species. 

The initial elemental abundances considered in our models were taken from \citet[][see the Solar values in their Table 1]{asplund09} except for Mg, Si, Cl, and F, which have been depleted by factors between 5.4 and 5700 (see Table \ref{atomabun} and references therein). For P, its elemental abundance has been depleted by a factor of 100 with respect to its Solar value (P/H=2.57$\times$10$^{-7}$), to simulate the fact that phosphorus is not detected in quiescent molecular clouds \citep{turner90}. Note that in the models of \citet{aota12}, a higher initial abundance of P/PH$_3$ of $\sim$10$^{-8}$ is assumed in the ices, which corresponds to a depletion factor of $\sim$10. However, a better agreement with the observations toward the B1 shocked region in the L1157 molecular outflow is obtained when this initial abundance is reduced to $\sim$3$\times$10$^{-9}$ (see their Section 3.2). This is also supported by recent findings by \citet{lefloch16} and \citet{rivilla16} in, respectively, L1157 B1 and a sample of massive star-forming regions. Therefore, it is reasonable to assume that P is likely depleted by a factor of 100 in molecular dark clouds.

\begin{deluxetable}{lcc}
\tablecaption{Assumed elemental abundances with respect to n$_H$.}
\tablewidth{0pt}
\tablehead{
\colhead{Element} & \colhead{Abundance} & \colhead{Ref.}
}
\startdata
		He & 0.085 & 1 \\
		C  & 2.692$\times$10$^{-4}$ & 1 \\
		O  & 4.898$\times$10$^{-4}$ & 1 \\
		N  & 6.761$\times$10$^{-5}$ & 1 \\
		S  & 1.318$\times$10$^{-5}$ & 1 \\
		Mg  & 7.0$\times$10$^{-9}$ & 2 \\
		Si  & 8.0$\times$10$^{-9}$ & 2 \\
		Cl  & 1.58$\times$10$^{-9}$ & 3 \\
		P  & 2.57$\times$10$^{-9}$ & 4 \\
		F  & 6.7$\times$10$^{-9}$ &  5 \\
\enddata
\tablenotetext{1}{Extracted from \citet{asplund09}.}
\tablenotetext{2}{As in the low-metal abundance case from \citet{graedel82} and \citet{morton74}.}
\tablenotetext{3}{Taken from \citet{schilke95}, \citet{cernicharo10} and \citet{codella12}.}
\tablenotetext{4}{As inferred by \citet{lefloch16}.}
\tablenotetext{5}{As derived by \citet{neufeld05}.}
\label{atomabun}
\end{deluxetable}

\uclchem is run in three phases. Phase 0 simulates the chemistry in a diffuse cloud with density n(H)=100 cm$^{-3}$ and temperature 20 K for 10$^6$ yrs. We have also tested higher temperatures for the Phase 0 stage \citep[T= 100 K; see][]{turner99} and the abundances of the P-bearing species at the end of Phase 1 (the collapse phase, see below) differ by less than 10\%. In Phase 1, the cloud undergoes free-fall collapse keeping the temperature constant at 10 K until the final density (n(H)=2$\times$10$^4$, 2$\times$10$^5$ and 2$\times$10$^6$ cm$^{-3}$) is reached. As explained in Section \ref{sec-collapse}, we have investigated the effects of a short-lived/long-lived collapse phase since it has been found to have important implications for the chemistry of PN and PO \citep[see e.g.][]{aota12,lefloch16}. Finally, Phase 2 simulates the physical processes associated with star formation such as warm-up of the protostellar envelope \citep[up to 50, 100 and 300 K, as described in][]{viti04}, interaction of C-type shocks \citep[for shock velocities of v$_s$=20 and 40 km s$^{-1}$ and using the parametrization of the physical structure of C-shock waves formulated by][]{jimenez08}, intense UV radiation \citep[with $\chi$=1, 10$^2$ and 10$^4$ Habing, as explored in][]{viti99} and cosmic rays \citep[with $\zeta$=1.3$\times$10$^{-17}$, 1.3$\times$10$^{-15}$ and 1.3$\times$10$^{-13}$ s$^{-1}$; see][]{harada15}. 

\begin{deluxetable}{lcccccc}
\tablecaption{Physical conditions for the collapse models.}
\tablewidth{0pt}
\tablehead{
\colhead{n(H)} & \colhead{r$_{\rm out}$} & \colhead{A$_v$} & \colhead{T} & \colhead{$\chi$} & \colhead{$\zeta$} & \colhead{Collapse\tablenotemark{1}} \\
\colhead{(cm$^{-3}$)} & (pc) & (mag) & \colhead{(K)} & \colhead{(Habing)} & \colhead{(s$^{-1}$)} & \colhead{}
}
\startdata
\multicolumn{7}{c}{Collapse Models} \\ \hline
2$\times$10$^4$ & 0.3 & 13 & 10 & 1 & 1.3$\times$10$^{-17}$ & long/short \\ 
2$\times$10$^5$ & 0.3 & 120 & 10 & 1 & 1.3$\times$10$^{-17}$ & long/short \\
2$\times$10$^6$ & 0.3 & 1200 & 10 & 1 & 1.3$\times$10$^{-17}$ & long/short \\ \hline
\multicolumn{7}{c}{Proto-stellar Heating Models} \\ \hline
2$\times$10$^4$ & 0.3 & 13 & 50 & 1 & 1.3$\times$10$^{-17}$ & long/short \\
2$\times$10$^4$ & 0.3 & 13 & 100 & 1 & 1.3$\times$10$^{-17}$ & long/short \\
2$\times$10$^4$ & 0.3 & 13 & 300 & 1 & 1.3$\times$10$^{-17}$ & long/short \\ 
2$\times$10$^5$ & 0.3 & 120 & 50 & 1 & 1.3$\times$10$^{-17}$ & long/short \\
2$\times$10$^5$ & 0.3 & 120 & 100 & 1 & 1.3$\times$10$^{-17}$ & long/short \\
2$\times$10$^5$ & 0.3 & 120 & 300 & 1 & 1.3$\times$10$^{-17}$ & long/short \\ 
2$\times$10$^6$ & 0.3 & 1200 & 50 & 1 & 1.3$\times$10$^{-17}$ & long/short \\
2$\times$10$^6$ & 0.3 & 1200 & 100 & 1 & 1.3$\times$10$^{-17}$ & long/short \\
2$\times$10$^6$ & 0.3 & 1200 & 300 & 1 & 1.3$\times$10$^{-17}$ & long/short \\ \hline
\multicolumn{7}{c}{UV Illumination Models} \\ \hline
2$\times$10$^4$ & 0.03 & 3 & 50 & 1 & 1.3$\times$10$^{-17}$ & long/short \\
2$\times$10$^4$ & 0.03 & 3 & 50 & 10$^2$ & 1.3$\times$10$^{-17}$ & long/short \\
2$\times$10$^4$ & 0.03 & 3 & 50 & 10$^4$ & 1.3$\times$10$^{-17}$ & long/short \\
2$\times$10$^4$ & 0.03 & 3 & 100 & 1 & 1.3$\times$10$^{-17}$ & long/short \\
2$\times$10$^4$ & 0.03 & 3 & 100 & 10$^2$ & 1.3$\times$10$^{-17}$ & long/short \\
2$\times$10$^4$ & 0.03 & 3 & 100 & 10$^4$ & 1.3$\times$10$^{-17}$ & long/short \\
2$\times$10$^4$ & 0.15 & 7.5 & 50 & 1 & 1.3$\times$10$^{-17}$ & long/short \\
2$\times$10$^4$ & 0.15 & 7.5 & 50 & 10$^2$ & 1.3$\times$10$^{-17}$ & long/short \\
2$\times$10$^4$ & 0.15 & 7.5 & 50 & 10$^4$ & 1.3$\times$10$^{-17}$ & long/short \\
2$\times$10$^4$ & 0.15 & 7.5 & 100 & 1 & 1.3$\times$10$^{-17}$ & long/short \\
2$\times$10$^4$ & 0.15 & 7.5 & 100 & 10$^2$ & 1.3$\times$10$^{-17}$ & long/short \\
2$\times$10$^4$ & 0.15 & 7.5 & 100 & 10$^4$ & 1.3$\times$10$^{-17}$ & long/short \\ 
2$\times$10$^4$ & 0.3 & 13 & 50 & 1 & 1.3$\times$10$^{-17}$ & long/short \\
2$\times$10$^4$ & 0.3 & 13 & 50 & 10$^2$ & 1.3$\times$10$^{-17}$ & long/short \\
2$\times$10$^4$ & 0.3 & 13 & 50 & 10$^4$ & 1.3$\times$10$^{-17}$ & long/short \\
2$\times$10$^4$ & 0.3 & 13 & 100 & 1 & 1.3$\times$10$^{-17}$ & long/short \\
2$\times$10$^4$ & 0.3 & 13 & 100 & 10$^2$ & 1.3$\times$10$^{-17}$ & long/short \\
2$\times$10$^4$ & 0.3 & 13 & 100 & 10$^4$ & 1.3$\times$10$^{-17}$ & long/short \\ \hline
\multicolumn{7}{c}{Enhanced CR Ionization Rate Models} \\ \hline
2$\times$10$^4$ & 0.3 & 13 & 10 & 1 & 1.3$\times$10$^{-15}$ & long/short \\
2$\times$10$^4$ & 0.3 & 13 & 10 & 1 & 1.3$\times$10$^{-14}$ & long/short \\
2$\times$10$^4$ & 0.3 & 13 & 10 & 1 & 1.3$\times$10$^{-13}$ & long/short \\
2$\times$10$^4$ & 0.3 & 13 & 100 & 1 & 1.3$\times$10$^{-15}$ & long/short \\
2$\times$10$^4$ & 0.3 & 13 & 100 & 1 & 1.3$\times$10$^{-14}$ & long/short \\
2$\times$10$^4$ & 0.3 & 13 & 100 & 1 & 1.3$\times$10$^{-13}$ & long/short \\
2$\times$10$^5$ & 0.3 & 120 & 10 & 1 & 1.3$\times$10$^{-15}$ & long/short \\
2$\times$10$^5$ & 0.3 & 120 & 10 & 1 & 1.3$\times$10$^{-13}$ & long/short \\
2$\times$10$^5$ & 0.3 & 120 & 100 & 1 & 1.3$\times$10$^{-15}$ & long/short \\
2$\times$10$^5$ & 0.3 & 120 & 100 & 1 & 1.3$\times$10$^{-13}$ & long/short \\
\enddata
\label{tab-collapse}
\tablenotetext{1}{This refers to the length of the collapse: short-lived if Phase 1 is stopped when the final density is reached, or long-lived if the collapse is some 10$^5$ yrs longer after the final density is reached.}
\end{deluxetable}

\section{Results}
\label{models}

\subsection{Effects of collapse on the chemistry of P-bearing species}
\label{sec-collapse}

\begin{figure*}
\begin{center}
\includegraphics[angle=270,width=1.0\textwidth]{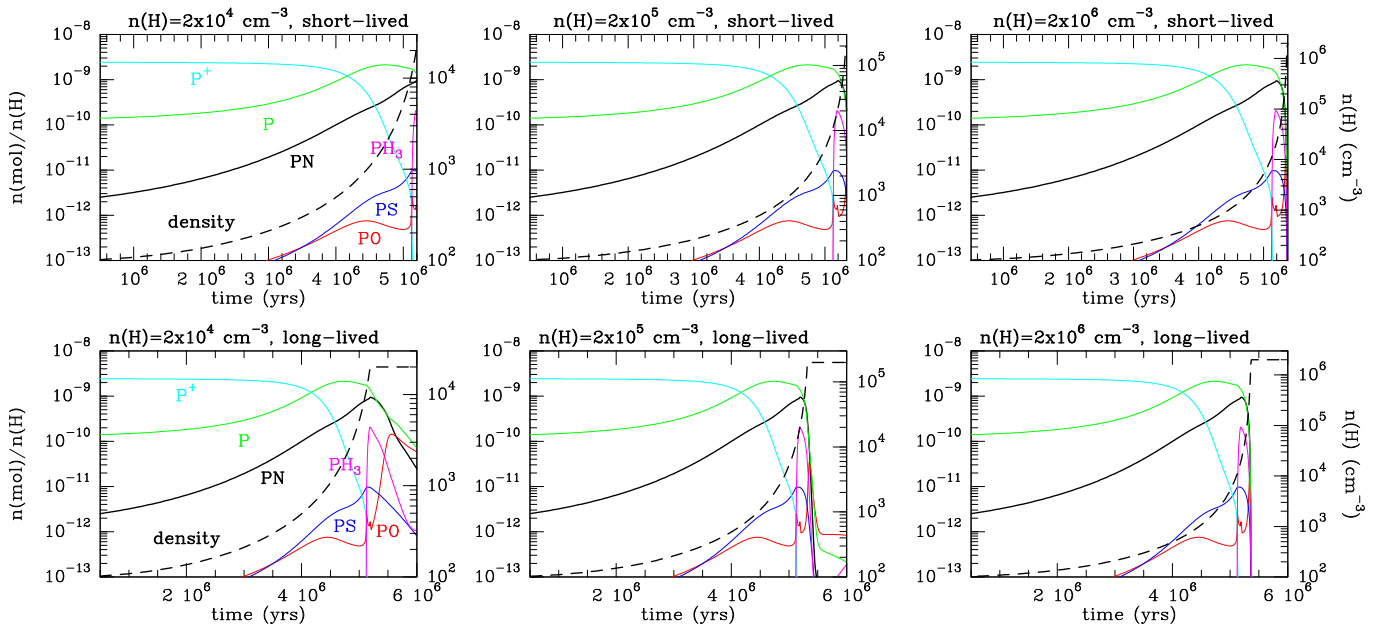}
\caption{Abundances of P-bearing species as a function of time simulated during the free-fall collapse of a cloud for final densities of n(H)=2$\times$10$^4$ (left), 2$\times$10$^5$ (middle), and 2$\times$10$^6$ cm$^{-3}$ (right). The length of the collapse is varied so that the chemistry stops when the final density is reached (short-lived collapse; upper panels) or the chemistry is let to evolve for a few 10$^5$ yrs more after the final density is reached (long-lived collapse; lower panels).}
\label{collapse}
\end{center}
\end{figure*}

In Figure \ref{collapse} we show the results obtained for the collapse phase (Phase 1 in \uclchem) for final gas densities n(H)=2$\times$10$^4$, 2$\times$10$^5$, and 2$\times$10$^6$ cm$^{-3}$ considering a short-lived collapse (the code stops once the final density is reached, i.e. at time-scales of 5.2-5.4$\times$10$^6$ yrs depending on the final density) and a long-lived collapse (chemistry runs until a final time-scale of 6$\times$10$^6$ yrs, i.e. between 6-8$\times$10$^5$ years longer after the final density is attained; see also Table \ref{tab-collapse}). In earlier works \citep[see][]{aota12,lefloch16}, the length of the collapse was found to have a strong impact on the subsequent evolution of the chemistry of PO and PN, and on their predicted abundance ratio. Indeed, in the short-lived scenario the gas-phase abundance of atomic nitrogen by the end of the collapse is large, which enhances the formation of PN and yields PO/PN ratios $\leq$1. On the contrary, if the collapse is long-lived, atomic nitrogen has enough time to deplete on grain surfaces and get converted mainly into NH$_3$, giving as a result PO/PN ratios $\geq$1 \citep{lefloch16}.  We therefore explore the consequences of the length of the collapse in our models by making the core become static for an extra 6-8$\times$10$^5$ years after the end of the collapse. This extra time is consistent with the estimated lower-limit dynamical age of pre-stellar cores \citep[$\geq$1-2$\times$10$^5$ years; see Section 4 in][]{pagani12} and provides a good match to observations toward this type of cores \citep{quenard18}.

In our models, we assume a physical radius for the core of 0.3 pc, which implies visual extinctions of 13 mag, 120 mag and 1200 mag for hydrogen densities of n(H)=2$\times$10$^4$, 2$\times$10$^5$ and 2$\times$10$^6$ cm $^{-3}$ respectively, according to the following expression \citep{bohlin78,frerking82}:

\begin{equation}
A_{\rm v} = A_{\rm v0} + \frac{n(H)\,\,\, \times\,\,\, r_{out}}{1.6\times10^{21}} \,\, \rm mag.
\end{equation}

\noindent
where r$_{out}$ is the radius of the core, n(H) is the number density of hydrogen nuclei, A$_{\rm v0}$ is the background extinction assumed to be 2 mag, and A$_{\rm v}$ is the total extinction within the core.

As shown in Figure \ref{collapse}, for all models the most abundant P-bearing species during the collapse are atomic phosphorus (P) and PN, with maximum abundances of $\sim$5$\times$10$^{-10}$-10$^{-9}$ at time-scales of $\sim$5$\times$10$^6$ yrs (i.e. when the final density is reached). The formation of PN in detectable abundances starts with the gas-phase reactions N + CP $\rightarrow$ PN + C and P + CN $\rightarrow$ PN + C, and proceeds through the reaction N + PO $\rightarrow$ PN + O in the gas phase by the end of the collapse. The formation of PO and PS is delayed with respect to PN and it occurs in the gas-phase via reactions O + PH $\rightarrow$ PO + H and HPS$^+$ + e- $\rightarrow$ PS + H respectively. PH$_3$ dramatically increases its gas-phase abundance by the end of the collapse, because it is efficiently formed on grain surfaces via hydrogenation and subsequently non-thermally desorbed. Note also the dramatic drop of P$^+$ with density as a result of the enhanced freeze out for ions.

At the highest densities in the collapse, the gas-phase abundances of all P-bearing species start to decrease as a result of freezing out onto dust grains. This depletion is more extreme in the long-lived case since species can freeze out for longer reaching gas-phase abundances even lower than 10$^{-13}$. Except for models with n(H)=2$\times$10$^6$ cm$^{-3}$ where all P-bearing species are fully frozen out in both the long-lived and short-lived cases, these differences in depletion by the end of the collapse have an impact on the PO/PN abundance ratio predicted during Phase 2 (see Sections \ref{sec-warm}, \ref{sec-UV}, \ref{sec-CRs} and \ref{sec-shock}). 


\subsection{Effects of proto-stellar heating}
\label{sec-warm}

\begin{figure*}
\begin{center}
\includegraphics[angle=270,width=1.0\textwidth]{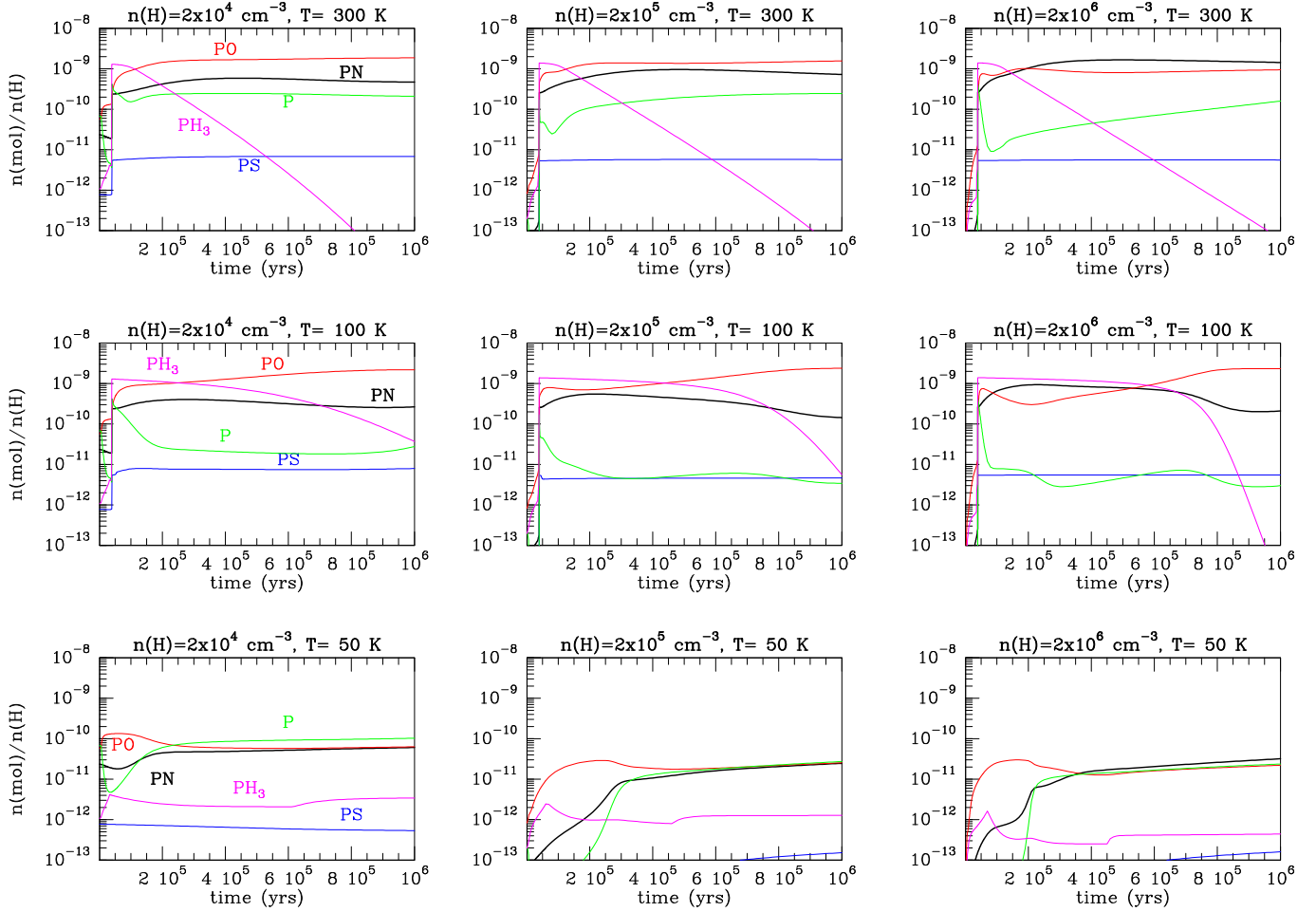}
\caption{Abundances of P-bearing species as a function of time simulated during the warming-up of the molecular envelope by a central protostar up to temperatures of 50 K (lower panels), 100 K (middle panels) and 300 K (upper panels). The final densities of the envelope are n(H)=2$\times$10$^4$ (left), 2$\times$10$^5$ (middle), and 2$\times$10$^6$ cm$^{-3}$ (right). These models have been run considering a long-lived collapse phase (Section \ref{sec-collapse}).}
\label{warmingup}
\end{center}
\end{figure*}

In this Section, we analyze the effects of the increase of dust and gas temperature due to the protostar on the chemistry of phosphorus. We consider a range of hydrogen gas densities (n(H)=2$\times$10$^4$, 2$\times$10$^5$, and 2$\times$10$^6$ cm$^{-3}$) and final temperatures (T = 50, 100 and 300 K) that are typically found in warm and hot cores (see Table \ref{tab-collapse}). We simulate the effect of the presence of a protostar at the center of the core by subjecting it to an increase in the gas and dust temperature. The temperature reaches its maximum value at the contraction time of the core (occurring at time-scales $\leq$10$^5$ yrs), following a power law derived by \citet{viti04} from the observational luminosity function of \citet{molinari00}. For simplicity, in our models we only consider the case of proto-stellar heating produced by a 15$\,$M$_\odot$ proto-star.

The results of the Phase 2 of our models are shown in Figure$\,$\ref{warmingup}. For T = 50 K, the only P-bearing that is released into the gas phase from the mantles is atomic P (at $\sim$3$\times$10$^4$ yrs). PH$_3$ is not thermally desorbed at such low temperatures and its solid abundance increases steadily during the warming-up phase thanks to hydrogenation (in particular, via $\#$PH$_2$ + $\#$H $\rightarrow$ $\#$PH$_3$, where $\#$ denotes a grain mantle species). However, once the temperature reaches its maximum and is kept constant to T = 50 K, hydrogenation ceases because atomic P is thermally desorbed at this temperature (see above) and PH$_3$ in the gas phase is slowly destroyed via proton transfer reactions with H$_3^+$, HCO$^+$ and H$_3$O$^+$. PO is formed in the gas phase via the reaction PH + O $\rightarrow$ PO + H while the formation of PN depends on the abundance of PO through N + PO $\rightarrow$ PN + O. In fact, these two species reach an equilibrium at time-scales $\sim$3-4$\times$10$^5$ yrs for all densities showing similar abundances of a few 10$^{-11}$. PS remains very low in all models with abundances $\leq$10$^{-12}$.

For T = 100 and 300 K, the evolution of the P-bearing species are similar presenting a first jump in their abundances at $\sim$5$\times$10$^4$yrs once the temperature of dust/gas reaches 100 K and all ices are thermally desorbed into the gas phase. Since PH$_3$ is the main reservoir of P in the mantles, this species is the most abundant for the first few 10$^5$ yrs. After that, it gets destroyed via proton transfer reactions with mainly H$_3$O$^+$. The decrease in the abundance of PH$_3$ is faster in models with T=300 K because the destruction reaction PH$_3$ + H $\rightarrow$ PH$_2$ + H$_2$ becomes efficient at higher temperatures due to its activation barrier of 735 K. Also, note that PH$_3$ is destroyed in the gas phase faster than other species such as PN and PO because, while PN/PO can re-form in the gas phase via reactions between P, PH and PH$_2$ and O, O$_2$ and N, the only gas-phase formation routes for PH$_3$ in our network are PH$_4^+$ + NH$_3$ $\rightarrow$ PH$_3$ + NH$_4^+$ and PH$_4^+$ + e$^-$ $\rightarrow$ PH$_3$ + H \citep[see also][]{thorne84}.

For PO, this molecule tends to increase its abundance with time in most models and, except for densities n(H)=2$\times$10$^6$ cm$^{-3}$, PO always stays above PN. 
Once PS is desorbed from the mantles of dust grains, its abundance remains practically constant at $\sim$5-6$\times$10$^{-12}$. In all these models the abundance of P$^+$ remains below 10$^{-13}$.

\begin{figure*}
\begin{center}
\includegraphics[angle=270,width=1.0\textwidth]{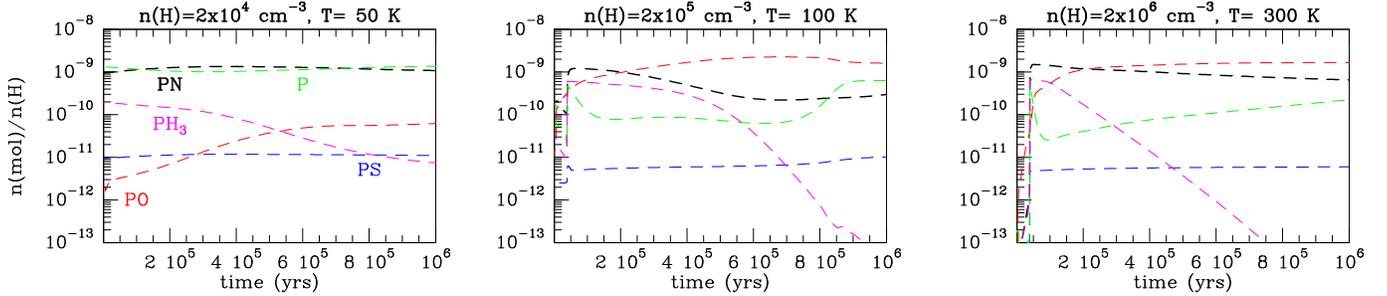}
\caption{Examples of the evolution of the abundances of P-bearing species as a function of time simulated during the warming-up of the molecular envelope by a central protostar considering a short-lived collapse phase. The abundances for models with n(H)=2$\times$10$^4$ cm$^{-3}$ and T=50 K are at least a factor of 10 higher than those for the long-lived collapse case. For higher temperatures, however, the behaviour of the P-bearing species is similar for both the short-lived and long-lived collapse scenarios.}
\label{warmingup-short}
\end{center}
\end{figure*}

The effects of the length of the collapse on the subsequent chemistry of P-bearing species during Phase 2 is illustrated in Figure \ref{warmingup-short}. For warm temperatures (T=50 K), the abundances in Phase 2 derived after considering a short-collapse phase, are enhanced by at least a factor of 10 with respect to the long-collapse phase (see left panel in Figure \ref{warmingup-short}). However, for higher temperatures, although the main P reservoir is PN instead of PH$_3$ at the beginning of Phase 2, the overall behaviour of all P-bearing species is similar to that of the long-collapse scenario since the gas phase chemistry clearly dominates.

\subsection{Effects of UV-photon radiation}
\label{sec-UV}

\begin{figure*}
\begin{center}
\includegraphics[angle=270,width=1.0\textwidth]{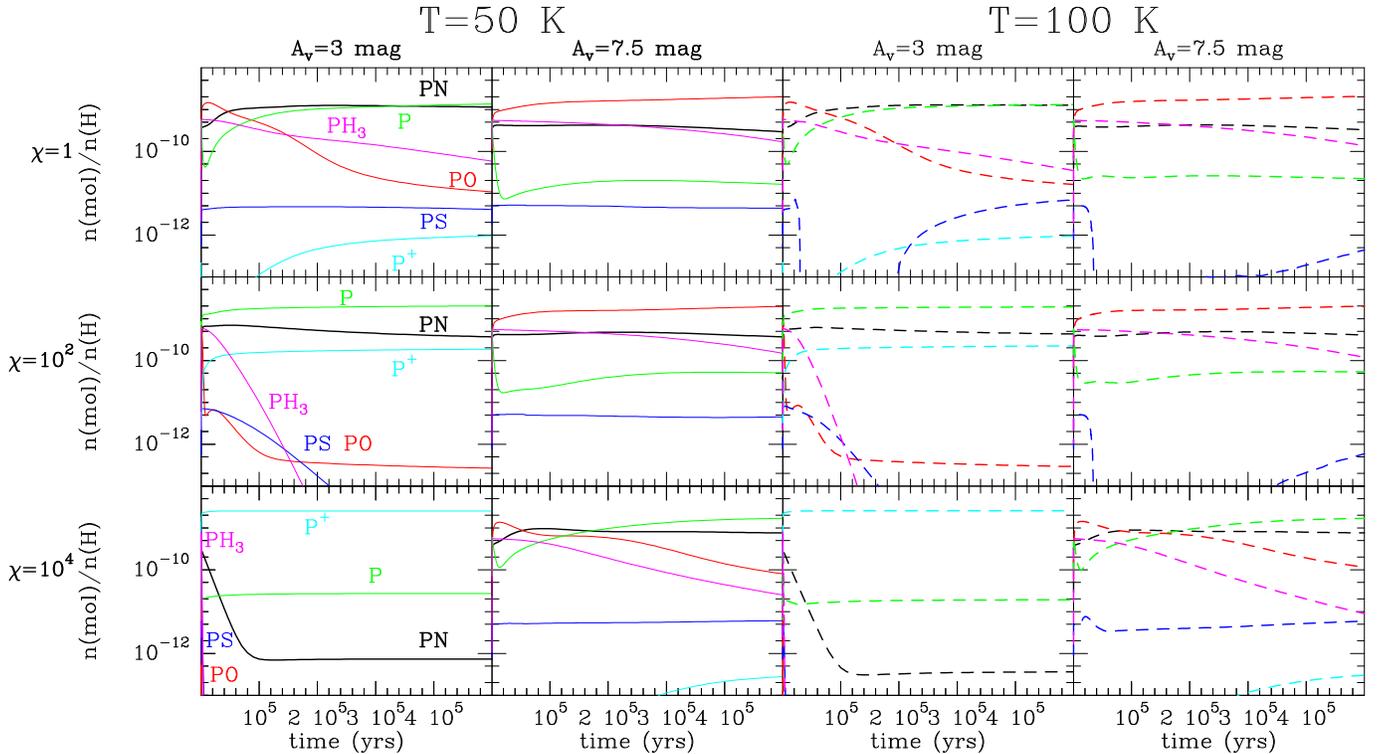}
\caption{Evolution of P-bearing species under the effects of a radiation field of $\chi$= 1, 100 and 10$^4$ Habing and for different visual extinctions (A$_v$=3 and 7.5 mag) after a long-lived collapse. Solid lines denote models with T= 50 K, while the results from models with T=100 K are shown in dashed lines. The results from models with A$_v$= 13 mag are identical to those obtained with A$_v$= 7.5 mag, and they are thus not shown.}
\label{UV-long}
\end{center}
\end{figure*}

\begin{figure*}
\begin{center}
\includegraphics[angle=270,width=0.7\textwidth]{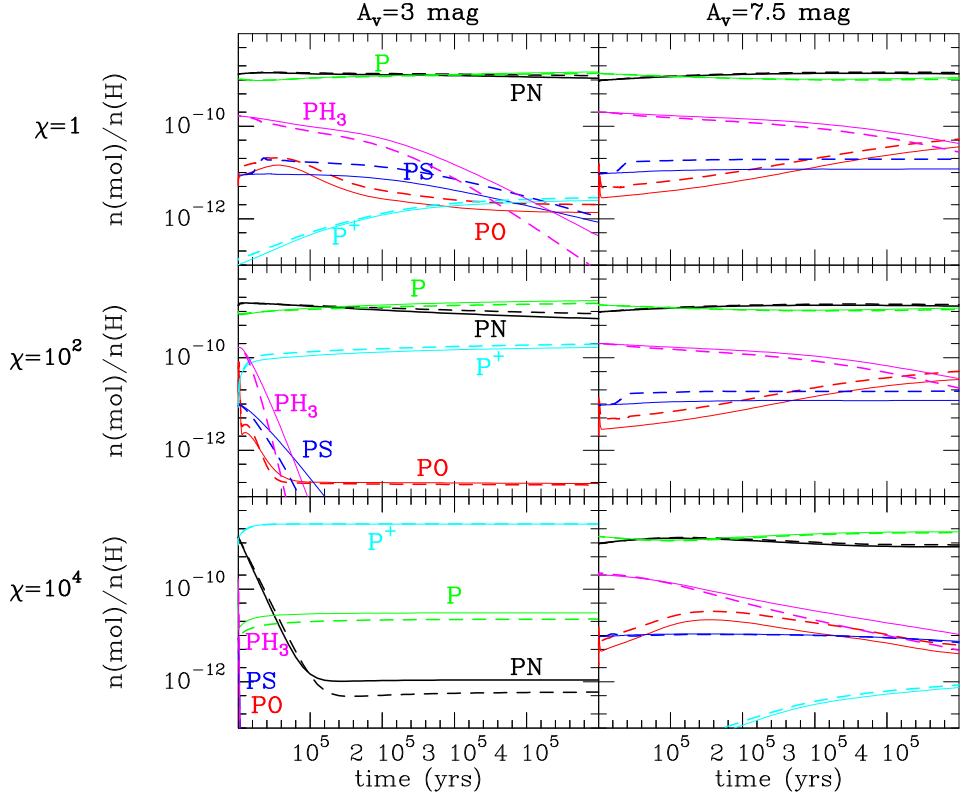}
\caption{The same as Figure \ref{UV-long}, but assuming a short-lived collapse phase. Solid lines indicate models with T= 50 K, dashed lines denote models with T=100 K. The abundance variations due to the change in temperature, are minimal. The results from models with A$_v$= 13 mag are identical to those obtained with A$_v$= 7.5 mag, and they are thus not shown.}
\label{UV-short}
\end{center}
\end{figure*}

The effects of intense UV illumination on the chemistry of phosphorus are evaluated by varying the interstellar radiation field within ranges typical of photon-dominated regions (PDRs). We consider UV-photon radiation fields of $\chi$= 100 and 10$^4$ Habing, visual extinctions of A$_{\rm v}$ = 3, 7.5 and 13 mag, and temperatures T = 50 and 100 K. These values are similar to those found in PDRs such as the Horsehead Nebula and the Orion Bar \citep{goicoechea09,schilke01}. In the models, the temperature is kept constant and the ices are instantaneously evaporated at the beginning of Phase 2. For completeness, we also present the results for models with $\chi$= 1 Habing (see Table \ref{tab-collapse}). The results are reported in Figures \ref{UV-long} and \ref{UV-short}. 

From Figure \ref{UV-long}, we find that at low extinctions (A$_{\rm v}$= 3 mag) PN and P are the most abundant P-bearing species (with abundances $\geq$5$\times$10$^{-10}$-10$^{-9}$) for low and intermediate UV radiation fields ($\chi$$\leq$100 Habing), while PO is efficiently destroyed. This is partly explained by the larger photo-destruction rate of PO (with $\alpha$=3$\times$10$^{-10}$ s$^{-1}$ and $\gamma$=2.0, where the rate is calculated as $k$=$\alpha$e$^{-\gamma \rm A_v}$) than for PN (with $\alpha$=5$\times$10$^{-12}$ s$^{-1}$ and $\gamma$=3.0). Note that these photo-destruction rates have been estimated from the ones for NO and N$_2$ \citep[see][]{harada10} and, therefore, may be inaccurate. However, even in the case that PN were destroyed at higher rates via UV photo-dissociation, it would re-form again in the gas phase thanks to the large amounts of atomic N and P available in the gas phase, which react via N + PO $\rightarrow$ PN + O and P + CN $\rightarrow$ PN + C yielding PN. Like PO, PH$_3$ and PS are also increasingly photo-dissociated with increasing UV radiation field. Note that for even higher values of the UV field ($\chi$= 10$^4$ Habing), all molecular material is photo-dissociated including PN, which shows abundances $\leq$10$^{-12}$.


For higher extinctions (A$_{\rm v}$=7.5 mag), the behaviour of all P-bearing species is similar to those obtained for the warming-up case with T= 50 and 100 K (see Figure \ref{warmingup}) since UV photo-dissociation becomes practically ineffective at A$_{\rm v}$$\geq$5 mag. As expected, the results of models with A$_{\rm v}$=13 (not shown in Figure \ref{UV-long}) are identical to those with A$_{\rm v}$=7.5 mag. The differences between models with T=50 K and T=100 K are minimal except for PS, whose chemistry is more sensitive to higher temperatures (see dashed lines in Figure \ref{UV-long}). 

In models with a short-lived collapse phase, PN is the most abundant P-bearing molecule with an abundance of $\sim$10$^{-9}$, following P closely. As for the long-lived case, only small differences are found between models with T= 50 K and T= 100 K. The results with A$_{\rm v}$= 7.5 mag are also very similar to the short-lived warm-up models from Figure \ref{warmingup-short}.   

We finally note that in all UV-photon illuminated models at low extinctions (A$_{\rm v}$=3 mag) or under high UV radiation fields ($\chi$= 10$^4$ Habing), the abundance of PN always remains above that of PO. 

\subsection{Effects of cosmic-rays}
\label{sec-CRs}

\begin{figure*}
\begin{center}
\includegraphics[angle=0,width=0.7\textwidth]{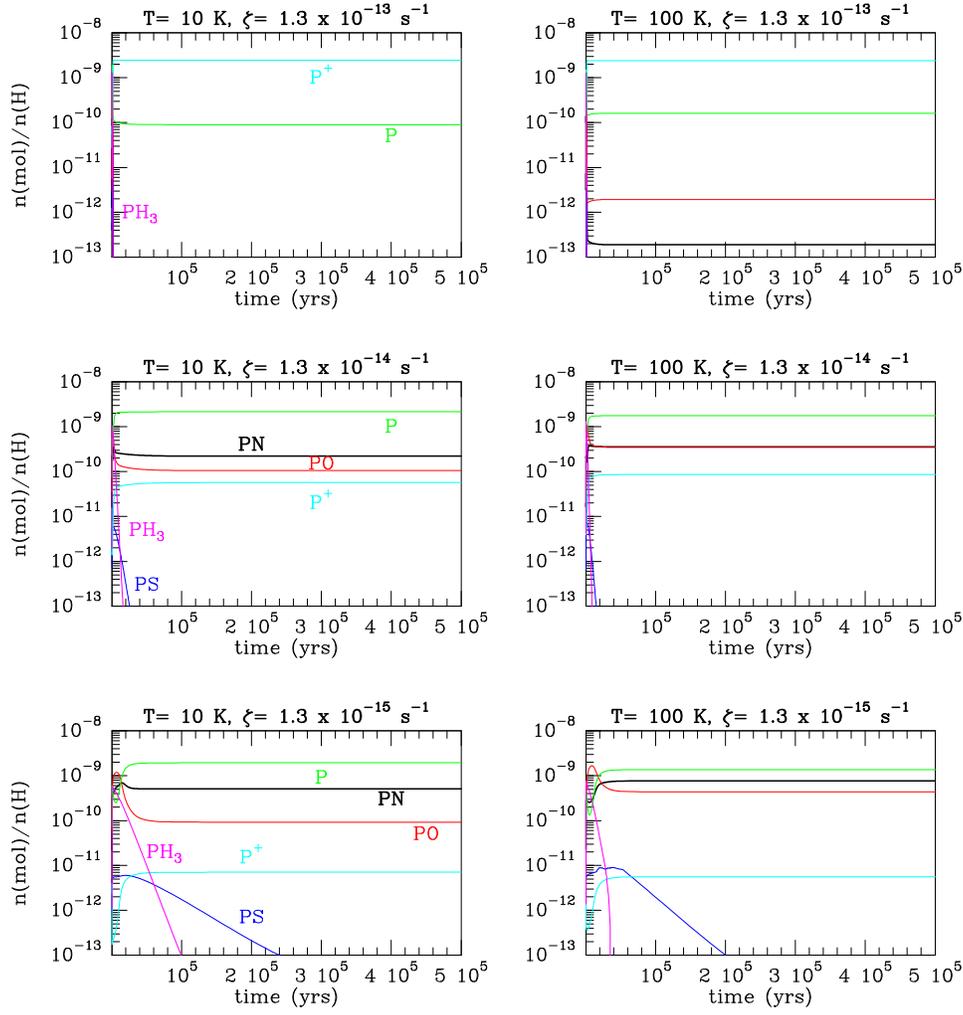}
\caption{Evolution of the abundances of P-bearing species as a function of time simulated for a hydrogen density n(H)=2$\times$10$^{4}$ cm$^{-3}$, enhanced cosmic-ray ionization rates $\zeta$ = 1.3$\times$10$^{-15}$, 1.3$\times$10$^{-14}$ and 1.3$\times$10$^{-13}$ s$^{-1}$, and gas temperatures T = 10 K and 100 K, considering a long-lived phase for the collapse.}
\label{CRs-1e4}
\end{center}
\end{figure*}

\begin{figure*}
\begin{center}
\includegraphics[angle=270,width=0.7\textwidth]{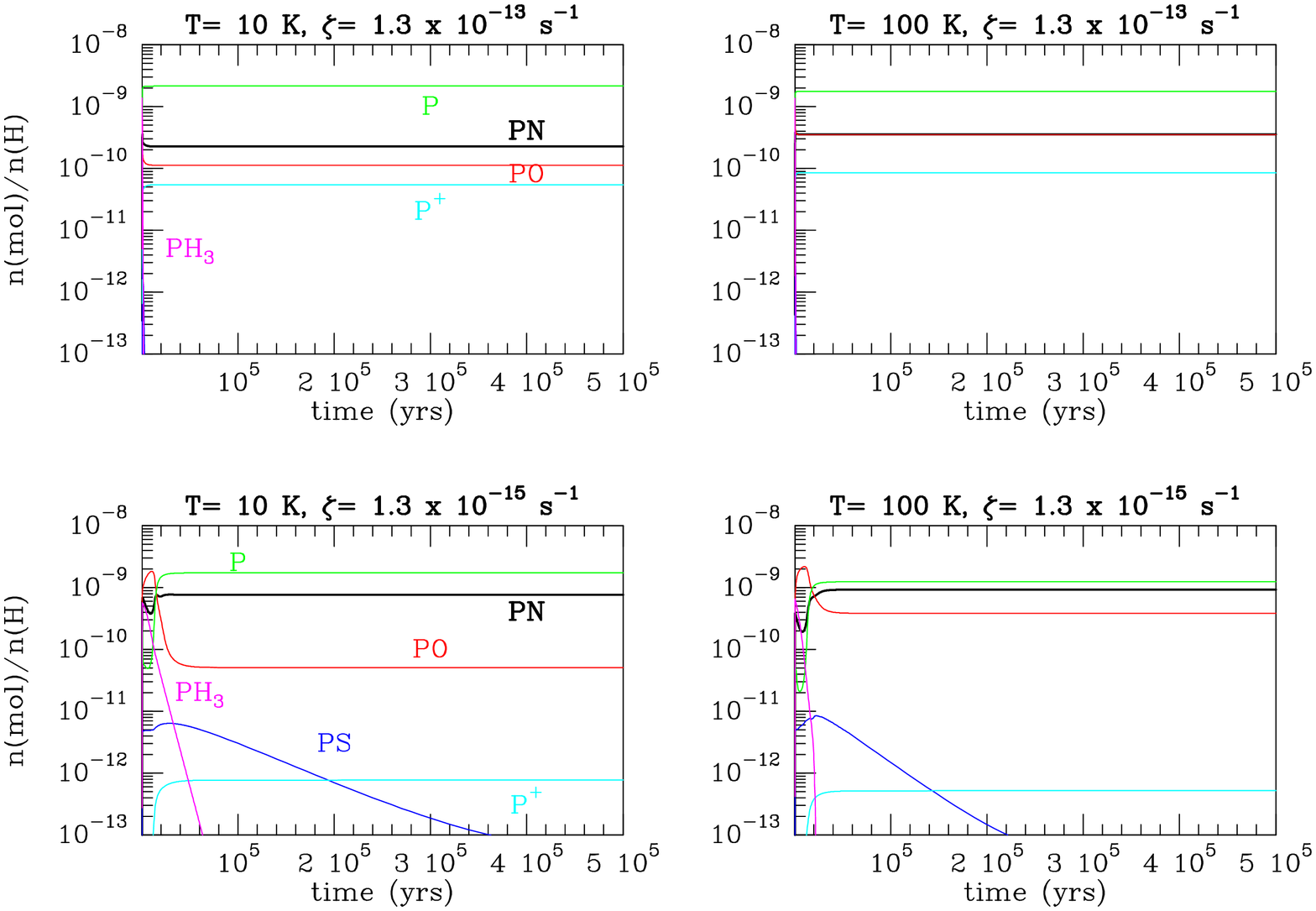}
\caption{Evolution of the abundances of P-bearing species as a function of time simulated for a hydrogen density n(H)=2$\times$10$^{5}$ cm$^{-3}$, enhanced cosmic-ray ionization rates $\zeta$ = 1.3$\times$10$^{-15}$ and 1.3$\times$10$^{-13}$ s$^{-1}$, and gas temperatures T = 10 K and 100 K, considering a long-lived phase for the collapse.}
\label{CRs-1e5}
\end{center}
\end{figure*}

We now evaluate the effects of cosmic-rays on the chemistry of P-bearing species. Cosmic-rays are present in a variety of environments from shocks in molecular outflows \citep{podio14}, to star-forming regions \citep{ceccarelli14,fontani17}, to the GMCs in the CMZ \citep{goto14}.
In Figures \ref{CRs-1e4} and \ref{CRs-1e5}, we present the evolution of the P-bearing species within a molecular cloud with hydrogen volume densities n(H)=2$\times$10$^4$ and 2$\times$10$^5$ cm$^{-3}$ and gas temperatures T = 10 and 100 K, under the influence of cosmic-rays whose ionization rates have been increased by factors 100, 1000 and 10$^4$ (i.e. $\zeta$=1.3$\times$10$^{-15}$, 1.3$\times$10$^{-14}$ and 1.3$\times$10$^{-13}$ s$^{-1}$, respectively).  For the T=100 K case we consider that the ices are instantaneously evaporated at the beginning of Phase 2, as in Section \ref{sec-UV}. The physical conditions assumed for these models are presented in Table \ref{tab-collapse}.

For the models with n(H)=2$\times$10$^4$ cm$^{-3}$ and T=10 K, Figure \ref{CRs-1e4} shows that a higher $\zeta$ (by factors 100 and 1000) enhances the abundances of species such as PN and PO with respect to those predicted by the collapse models at the end of the collapse (see lowermost left panel of Figure \ref{collapse}). This is due to i) the non-thermal desorption of these species by cosmic-ray induced secondary UV photons, and to ii) the rapid photo-dissociation of PH$_3$ by this secondary UV field. As already noted in Section \ref{sec-warm}, once PH$_3$ is destroyed, it cannot re-form efficiently in the gas phase \citep[][]{thorne84}, which unlocks high abundances of gas-phase P that can then be used to produce PN and PO. 

For $\zeta$=1.3$\times$10$^{-13}$ s$^{-1}$ (i.e. an enhancement in $\zeta$ by a factor of 10$^4$), molecules are rapidly photo-dissociated by the strong secondary UV-photon radiation field, yielding large abundances of P$^+$ and P. A similar trend is seen for higher temperatures (T=100 K), where PH$_3$ is rapidly destroyed in the gas phase in favour of PN and PO. In all models, PS is either destroyed or its abundance remains below $\leq$10$^{-12}$. PN lies above PO with PO/PN ratios $\sim$0.5-1 except for the case with T=100 K and $\zeta$=1.3$\times$10$^{-13}$ s$^{-1}$ where PO is a factor of $\sim$10 more abundant than PN.

For higher densities (n(H)=2$\times$10$^5$ cm$^{-3}$; see Figure \ref{CRs-1e5}), PN and PO are largely enhanced for $\zeta$=1.3$\times$10$^{-13}$ s$^{-1}$ with respect to the low-density case thanks to the higher abundance of atomic N and O in these models as they are more protected by the higher extinction. Atomic N and O yield higher abundances of PN and PO via reactions N + PO $\rightarrow$ PN + O and O + PH $\rightarrow$ PO + H, respectively. PN also tends to be more abundant than PO in the high-density models.

We note that all models above consider a long-lived collapse phase (see Section \ref{sec-collapse}). If the collapse phase is short-lived, the largest discrepancies are generally found for models with $\zeta$$\geq$1.3$\times$10$^{-14}$ s$^{-1}$, where the PN and PO abundances decrease by factors 10-100 with respect to the long-lived collapse case (see Figures \ref{CRs-1e4-short} and \ref{CRs-1e5-short}). This is explained by the fact that in the short-lived collapse, the ice reservoir of P-bearing molecules is not as large as in the long-lived case, and the gas-phase P-bearing species are rapidly destroyed by the cosmic-ray induced secondary UV photons. 

\subsection{Effects of C-type shock waves}
\label{sec-shock}

\subsubsection{General behaviour of P-bearing species in C-shocks}
\label{general}

\begin{deluxetable}{lccccc}
\tablecaption{Physical parameters assumed for the C-type shock models. All models assume $\chi$= 1 Habing.}
\tablewidth{0pt}
\tablehead{\colhead{n(H)} & \colhead{v$_{\rm s}$} & \colhead{T$_{\rm n,max}$} & \colhead{t$_{\rm sat}$} & \colhead{$\zeta$} & \colhead{Collapse\tablenotemark{1}} \\
\colhead{(cm$^{-3}$)} & \colhead{(km s$^{-1}$)} & \colhead{(K)} & \colhead{(years)} & \colhead{(s$^{-1}$)} & \colhead{}
}
\startdata
2$\times$10$^4$ & 20 & 900 & 57.0 & 1.3$\times$10$^{-17}$ & long/short \\
2$\times$10$^4$ & 20 & 900 & 57.0 & 1.3$\times$10$^{-15}$ & long/short \\
2$\times$10$^4$ & 20 & 900 & 57.0 & 1.3$\times$10$^{-13}$ & long/short \\
2$\times$10$^4$ & 40 & 2200 & 45.5 & 
1.3$\times$10$^{-17}$ & long/short \\
2$\times$10$^4$ & 40 & 2200 & 45.5 & 1.3$\times$10$^{-15}$ & long/short \\
2$\times$10$^4$ & 40 & 2200 & 45.5 &  1.3$\times$10$^{-13}$ & long/short \\ \hline
2$\times$10$^5$ & 20 & 800 & 5.7 & 1.3$\times$10$^{-17}$ & long/short \\
2$\times$10$^5$ & 20 & 800 & 5.7 & 1.3$\times$10$^{-15}$ & long/short \\
2$\times$10$^5$ & 20 & 800 & 5.7 & 1.3$\times$10$^{-13}$ & long/short \\
2$\times$10$^5$ & 40 & 4000 & 4.6 & 1.3$\times$10$^{-17}$ & long/short \\ 
2$\times$10$^5$ & 40 & 4000 & 4.6 & 1.3$\times$10$^{-15}$ & long/short \\
2$\times$10$^5$ & 40 & 4000 & 4.6 &  1.3$\times$10$^{-13}$ & long/short \\
\enddata
\label{tab-shocks}
\tablenotetext{1}{This refers to the length of the collapse during Phase 1 in our modelling (see also the caption in Table \ref{tab-collapse}).}
\end{deluxetable}

P-bearing species such as PN and PO have recently been observed in shocked regions in molecular outflows \citep{yamaguchi11,lefloch16}. In addition, these species have also been detected in the turbulent GMCs in the Central Molecular Zone \citep{rivilla18}, believed to be affected by widespread low-velocity shocks \citep{requena-torres06} and likely under the influence of enhanced cosmic-ray ionization rates \citep{goto14,harada15}. 

In Table \ref{tab-shocks} we present the physical parameters assumed for the C-type shock models used in our calculations. The pre-shock density is denoted by n(H), v$_s$ is the shock speed, T$_{n,max}$ refers to the maximum temperature of the neutral fluid attained within the shock and t$_{sat}$ is the saturation time at which most of the material within the icy mantles of dust grains is sputtered into the gas phase \citep[for more details, see][]{jimenez08,holdship17}. Different values of the cosmic-ray ionization rate ($\zeta$=10$^{-17}$, 10$^{-15}$, and 10$^{-13}$ s$^{-1}$) are also considered.

\begin{figure*}
\begin{center}
\includegraphics[angle=270,width=1.0\textwidth]{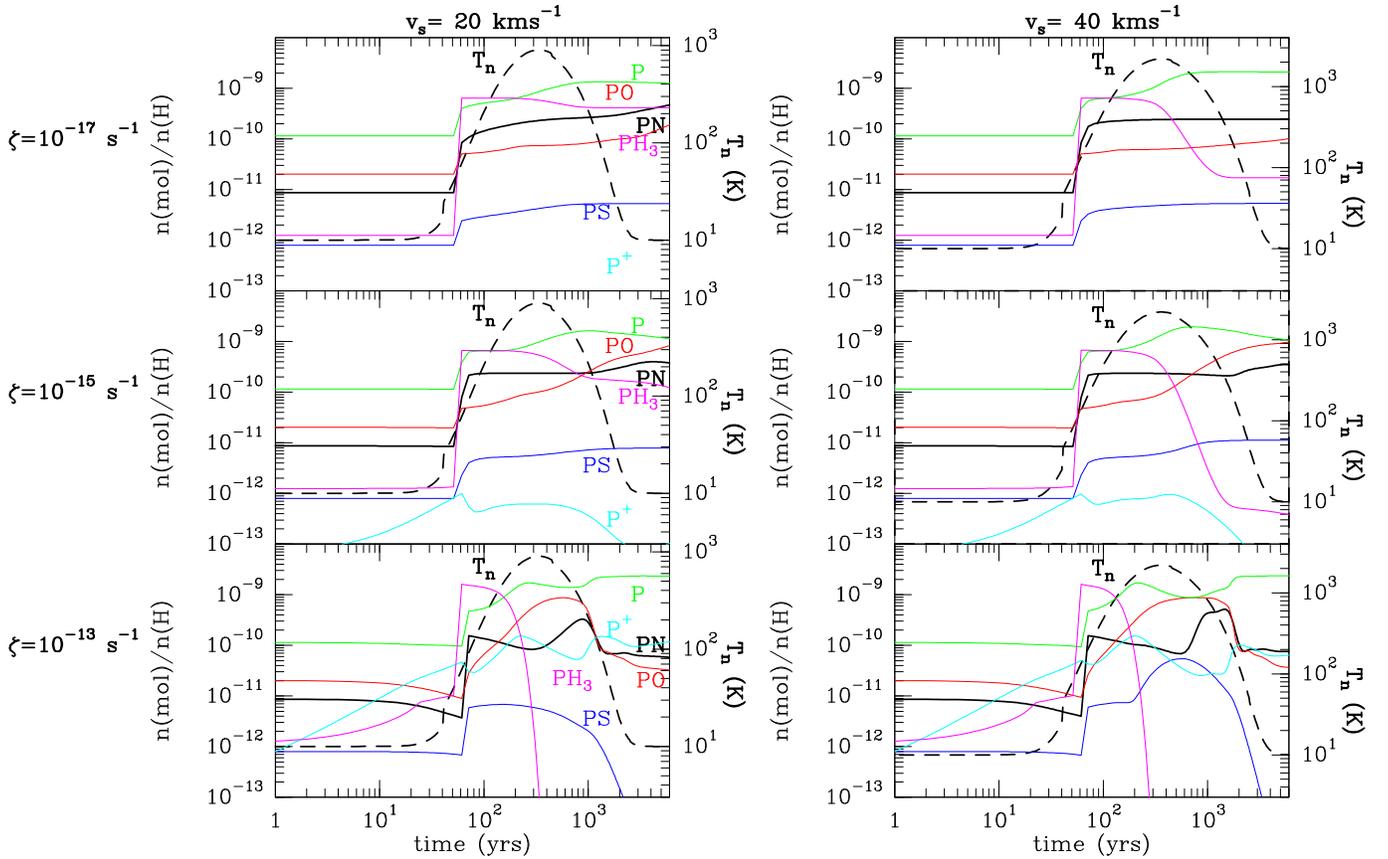}
\caption{Evolution of the abundances of P-bearing species as a function of time across two C-type shocks with a pre-shock density of n(H)=2$\times$10$^4$ cm$^{-3}$ and shock speeds of v$_s$=20 km s$^{-1}$ (left panels) and v$_s$=40 km s$^{-1}$ (right panels). Dashed lines indicate the evolution of the temperature of the neutral fluid within the shock (T$_n$). To simulate the extreme conditions in the Galactic Center, we also consider that the shocked gas is affected by enhanced cosmic ray ionization rates of $\zeta$= 10$^{-15}$ and 10$^{-13}$ s$^{-1}$. These models have been obtained considering a long-lived collapse.}
    \label{shocks-1e4}
\end{center}
\end{figure*}

\begin{figure*}
\begin{center}
\includegraphics[angle=270,width=1.0\textwidth]{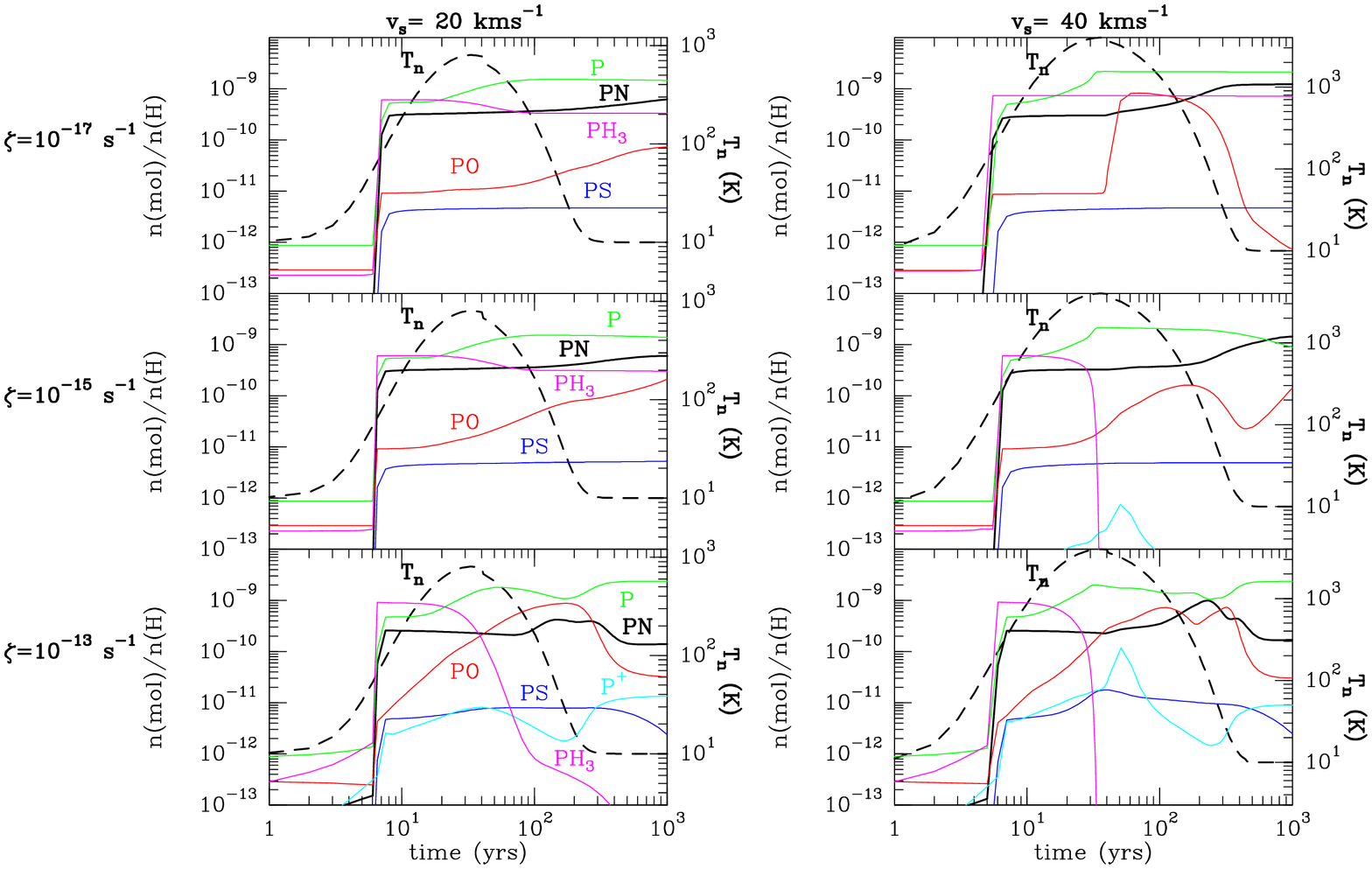}
\caption{The same as Figure \ref{shocks-1e5}, but for a pre-shock density of n(H)=2$\times$10$^5$ cm$^{-3}$.}
    \label{shocks-1e5}
\end{center}
\end{figure*}

Figures \ref{shocks-1e4} and \ref{shocks-1e5} show the evolution of P-bearing molecules across the C-type shock structure for hydrogen gas densities of n(H)=2$\times$10$^4$ cm$^{-3}$ and n(H)=2$\times$10$^5$ cm$^{-3}$, respectively. All species are enhanced early in the shock (at $\sim$60 yrs for n(H)=2$\times$10$^4$ cm$^{-3}$ and at $\sim$6 yrs for n(H)=2$\times$10$^5$ cm$^{-3}$) due to sputtering. In particular, PH$_3$ and PN present enhancements larger than factors 100 and 10, respectively. For models with 
shock speeds v$_s$=20 km s$^{-1}$, PH$_3$ stays relatively abundant throughout the shock after the sputtering of the ices (abundance between 10$^{-10}$-10$^{-9}$), except in the presence of a high cosmic-ray ionization rate when PH$_3$ is more efficiently destroyed thanks to reactions with He$^+$ and H$_3^+$ (see model with $\zeta$=1.3$\times$10$^{-13}$ s$^{-1}$ in Figure$\,$\ref{shocks-1e4}). For higher shock velocities (v$_s$= 40 km s$^{-1}$), PH$_3$ is destroyed due to the endothermic reaction PH$_3$ + H $\rightarrow$ PH$_2$ + H$_2$, whose activation energy is 735 K \citep[see Table 1 in][]{charnley94}. The destruction of PH$_3$ is even faster at higher temperatures (as in models with n(H)=2$\times$10$^5$ cm$^{-3}$ and v$_s$=40 km s$^{-1}$) or under the effects of cosmic rays. 

From Figures \ref{shocks-1e4} and \ref{shocks-1e5}, we find that PN remains relatively constant throughout the shock with abundances typically falling between 1-5$\times$10$^{-10}$. In contrast to PN, PO shows a different behaviour depending on the model, becoming more abundant than PN in shocks with n(H)=2$\times$10$^{4}$ cm$^{-3}$ and $\zeta$$\geq$10$^{-15}$ s$^{-1}$, or in shocks with n(H)=2$\times$10$^{5}$ cm$^{-3}$ and $\zeta$$\geq$10$^{-13}$ s$^{-1}$. PS experiences little changes in the gas phase after its injection into the gas phase by sputtering, except in models with $\zeta$=1.3$\times$10$^{-13}$ s$^{-1}$ where it is destroyed by H$_3^+$ and He$^+$.

For the short-lived collapse models, Figures \ref{shocks-1e4-short} and \ref{shocks-1e5-short} show that the most abundant P-bearing species within the shock are atomic P and PN, since they have not had enough time to deplete onto dust grains and to subsequently hydrogenate during the collapse phase. PO becomes comparable in abundance to PN only for high cosmic-ray ionization rates ($\zeta$$\sim$10$^{-13}$ s$^{-1}$). 

\subsubsection{Comparison with the \citet{lefloch16}'s models}
\label{compar-lefloch}

\citet{lefloch16} also presented models of the chemistry of P-bearing species in C-type shock waves using \texttt{uclchem} and the C-shock parametric approximation of \citet{jimenez08}. The chemical network used in these models, however, did not include either PH$_3$ or the endothermic reactions PH$_3$ + H $\rightarrow$ PH$_2$ + H$_2$, PH$_2$ + H $\rightarrow$ PH + H$_2$, and PH + H $\rightarrow$ P + H$_2$ \citep[with energy barriers ranging from 318 to 735 K;][]{charnley94}. This has important consequences in the evolution of the molecular abundances of PO and PN and its ratio in the shock since, as opposed to the results of \citet{lefloch16}, PN is always more abundant than PO in our models with standard values of the cosmic-ray ionization rate (i.e. $\zeta$=10$^{-17}$ s$^{-1}$; see Figures \ref{shocks-1e4} and \ref{shocks-1e5}). If the exothermic reactions of PH$_3$, PH$_2$ and PH with H are {\it switched-off} in our models, we recover the results of \citet{lefloch16}.

Millimeter observations of PN and PO in outflows \citep{lefloch16} and in Galactic Center Giant Molecular Clouds \citep{rivilla18}, however, indicate that the PO/PN abundance ratio is always $\geq$1, which can be reconciled with our models only if i) cosmic rays are significantly enhanced; or ii) the chemical network of PO is incomplete. In the following section (Section \ref{missing}), we explore the possibility that the chemical network of PO lacks a few key formation reactions.

\section{New formation reaction for PO}
\label{missing}

\begin{deluxetable}{cccc}
\tablecaption{Reaction rate assumed for the models of Section \ref{missing}, as extracted from UMIST.}
\tablewidth{0pt}
\tablehead{\colhead{Reaction)} & \colhead{$\alpha$ (cm$^{3}$ s$^{-1}$)} & \colhead{$\beta$} & \colhead{$\gamma$ (K)} 
}
\startdata
P + OH $\rightarrow$ PO + H & 6.1$\times$10$^{-11}$ & -0.23 & 14.9 \\
\enddata
\label{tab-ratesPO}
\end{deluxetable}

\citet{codella18} have recently carried out the modelling of the chemistry of NO (a species with a molecular structure similar to that to PO) in hot corinos and shocked gas in molecular outflows. These authors have noted that the enhancement of NO in the L1157-B1 shocked region, is mainly produced by the reaction of atomic nitrogen, N, with OH. The equivalent reaction for P (P + OH $\rightarrow$ PO + H), however, does not exist in any database and, given the similarity between NO and PO, the reaction between P and OH may also be efficient (note that the activation barrier is only of $\sim$15 K; see Table \ref{tab-ratesPO}). 

\begin{figure*}
\begin{center}
\includegraphics[angle=270,width=0.8\textwidth]{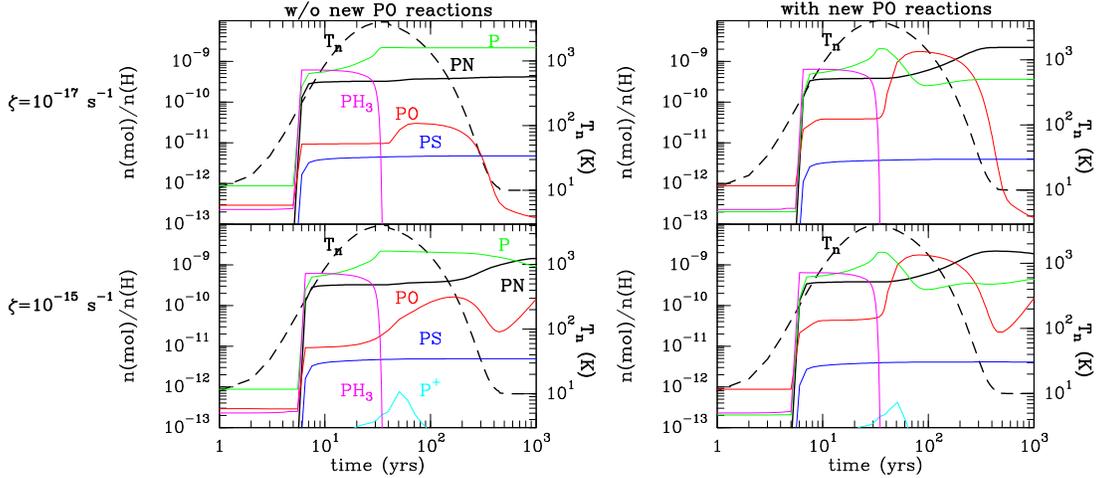}
\caption{Comparison of models obtained for n(H)=2$\times$10$^5$ cm$^{-3}$ and v$_s$=40 km s$^{-1}$ using the original chemical network (as presented in Section \ref{network}; left panels) and a new network where the proposed PO formation route P + OH $\rightarrow$ PO + H is included. Models with comic-ray ionization rates of $\zeta$$\sim$10$^{-17}$ s$^{-1}$ (upper panels) and $\zeta$$\sim$10$^{-15}$ s$^{-1}$ (lower panels) are presented. All models shown here consider a long-lived collapse phase.}
    \label{lefloch}
\end{center}
\end{figure*}

We have therefore re-run all our models after including the reaction P + OH $\rightarrow$ PO + H in our chemical network. As reaction rate, we have assumed the one from the analogous reaction with NO (see Table \ref{tab-ratesPO}). Our results show that for most models the inclusion of this reaction does not affect appreciably the final abundance of PO, since it is enhanced only by factors $\sim$2-3. However, for models with C-type shock waves, the abundance of PO can be enhanced by several orders of magnitude (see Figure \ref{lefloch}). From this Figure, it is clear that thanks to the new reaction P + OH $\rightarrow$ PO + H, PO even becomes more abundant than PN within the shock for standard cosmic-ray ionization rates, in contrast to our results from Section \ref{sec-shock}. This occurs thanks to the high temperatures (of thousands of K) attained within the shock. In fact, if this reaction were as efficient as proposed here, the PO/PN ratio in shocks would be dominated by this reaction and not by enhanced cosmic rays any longer (see lower panels of Figure \ref{lefloch}). Theoretical calculations and/or experiments are needed to clearly establish the efficiency of the PO gas-phase formation route P + OH $\rightarrow$ PO + H.

\section{Comparison with observations}
\label{sec-comparison-obs}

\subsection{PN and PO}

In this section we compare our modelling results to the abundances of PN and PO measured toward low-mass/high-mass starless/pre-stellar cores \citep{turner90,mininni18}, massive hot cores \citep{turner90,rivilla16}, GMCs in the Galactic Center \citep{rivilla18}, and molecular outflows \citep{yamaguchi11,lefloch16}: 

\begin{itemize}
\item {\it Cold Cloud and Starless/Pre-stellar Cores}: Early millimeter observations of PN toward a sample of low-mass cold cloud cores (with T$\sim$10 K) provided upper limits to the abundance of PN of $\leq$0.2-4$\times$10$^{-11}$ in cold cloud cores \citep{turner90}. Recent higher-sensitivity observations performed toward the L1544 pre-stellar core reveal that PN remains undetected with upper limits of $\leq$4.6$\times$10$^{-13}$ \citep[as derived from the dataset of][]{jimenez16}. As shown in Figure \ref{collapse} for the long-lived collapse case (see bottom panels), these upper limits are consistent with the low gas-phase PN abundances predicted at a few 10$^5$ yrs after the end of the collapse in cold high-density cores (with hydrogen gas densities $\geq$2$\times$10$^5$ cm$^{-3}$). 
For massive starless cores, which show slightly warmer temperatures \citep[with T$\sim$25-30 K;][]{fontani16}, the derived abundances of PN are $\sim$10$^{-11}$ and 5$\times$10$^{-12}$ \citep{mininni18}, in agreement with those predicted by our models under warm temperatures for n(H)$\geq$2$\times$10$^5$ cm$^{-3}$ and time-scales of a few 10$^5$ yrs (see models with T=50 K in Figure \ref{warmingup}). No upper limits are available for PO toward low-mass/high-mass cold cores. 

\item {\it Massive Hot cores}: The derived abundance of PN in hot molecular cores typically range from some 10$^{-12}$ to almost 10$^{-9}$ \citep{turner90,rivilla16}. The gas temperatures and hydrogen gas densities derived for these regions are $\sim$30-150 K and $\sim$10$^5$-10$^6$ cm$^{-3}$ respectively \citep[see e.g.][]{turner90}. The PN abundances predicted for time-scales $\leq$2-3$\times$10$^5$ yrs (typical of hot cores) by models with T$\sim$50-100 K and n(H)$\sim$2$\times$10$^5$-2$\times$10$^6$ cm$^{-3}$ provide similar PN abundances to those observed (see upper and middle panels in Figure \ref{warmingup}). The PO abundances obtained in our models ($\sim$0.3-6.0$\times$10$^{-10}$) also reproduce well those observed in hot molecular cores \citep{rivilla16}.

\item {\it Galactic Center GMCs}: 
\citet{rivilla18} have recently carried out observations of PN and PO toward a sample of quiescent GMCs in the Galactic Center that present either shock-dominated or radiation-dominated chemistries. \citet{rivilla18} found a positive trend between the abundances of PN and $^{29}$SiO, which suggests that PN may be formed in shocks as SiO. The derived PN abundances toward shock-dominated, quiescent GMCs are $\sim$0.5-5$\times$10$^{-11}$ \citep{rivilla18}, i.e. within a factor of 10 those predicted by our models with v$_s$=20 km s$^{-1}$, n(H)=2$\times$10$^4$ cm$^{-3}$ and high cosmic ray ionization rates (see Figure \ref{shocks-1e4}, left panels). The measured PO/PN ratio of 1.5 in the quiescent GMC G+0.693-0.03 also supports the models with enhanced cosmic rays, although, as discussed in Section \ref{missing}, a PO/PN ratio $\geq$1 could also be obtained in shocks if the reaction P + OH $\rightarrow$ PO + H were efficient. This reaction is currently not included in any chemical database.
For the quiescent GMCs affected by radiation, only upper limits to the abundance of PN are available. This is consistent with our results of the chemistry of P-bearing species in PDRs, for which UV radiation fields $\chi$$\sim$10$^4$ Habing destroy PN and PO down to abundances $\leq$10$^{-12}$ (for A$_v$=3 mag; see Figures \ref{UV-long} and \ref{UV-short}).  The lack of detection of PN in the Orion Bar \citep[][]{cuadrado15,rivilla18}, confirms this hypothesis \citep[the Orion Bar is affected by a UV radiation field of $\chi$=5$\times$10$^4$ Habing; see e.g.][]{schilke01}. 

\item {\it Molecular outflows}:
In \citet{lefloch16}, the preferred model to explain the PO/PN abundance ratio of $\sim$2 measured in the L1157-B1 shock cavity, involves a shock with gas densities of 10$^5$ cm$^{-3}$, shock speeds of 40 km s$^{-1}$, and a standard cosmic-ray ionization rate (1.5$\times$10$^{-17}$ s$^{-1}$). Unless the chemical network of PO is incomplete (see Section \ref{missing}), our shock model results suggest that the chemistry of P-bearing species in L1157-B1 is likely affected by cosmic rays. This is consistent with the detection of molecular ions such as HOCO$^+$ and SO$^+$ in L1157-B1, which requires a cosmic-ray ionization rate of $\sim$3$\times$10$^{-16}$ s$^{-1}$ \citep{podio14}. Theoretical studies and/or new experiments are needed to establish whether the reaction P + OH $\rightarrow$ PO + H is indeed as efficient as proposed in Table \ref{tab-ratesPO}.

\end{itemize}

\subsection{Phosphine (PH$_3$)}

Although phosphine (PH$_3$) has been detected in the atmospheres of Jupiter and Saturn \citep{larson77,weisstein94}, there is no evidence of its presence in star-forming regions. 
According to our models in Section \ref{models}, PH$_3$ tends to be destroyed rapidly in the gas phase (for time-scales shorter than a few 10$^4$ years) after its ejection from dust grains, as also noted by \citet{charnley94}. Nevertheless, for some cases PH$_3$ could remain in the gas phase for longer time-scales in hot molecular cores and in shocked regions, reaching gas-phase abundances as high as $\sim$10$^{-9}$ (see Figures \ref{warmingup}, \ref{shocks-1e4} and \ref{shocks-1e5}). PH$_3$ could then be observed via its $J$=1$\rightarrow$0 and $J$=2$\rightarrow$1 rotational transitions at 266.9 and 533.8 GHz respectively, with instruments such as the IRAM 30m telescope, ALMA and SOFIA. To our knowledge, no observations (or reported upper limits) do exist for this molecule toward hot molecular cores. However, upper limits to the abundance of PH$_3$ have been reported toward the B1 shocked region in the L1157 molecular outflow \citep[of $\leq$10$^{-9}$;][]{lefloch16}. These upper limits, which were inferred from high-sensitivity IRAM 30m telescope data\footnote{They belong to the ASAI Large Program (Astrochemical Surveys at IRAM), and the upper limits of the PH$_3$ abundance were estimated using a rms noise level of 3 mK over a linewidth of 5 km s$^{-1}$.}, are consistent with the abundances predicted by our models considering the interaction of shock waves (Figures \ref{shocks-1e4} and \ref{shocks-1e5}).

\section{The PO/PN ratio}
\label{sec-ratio}

For the sources where PO has been detected in the ISM \citep[e.g. hot cores, molecular outflows and the quiescent Galactic Center GMC G+0.693 in the Galactic Center; see][]{rivilla16,lefloch16,rivilla18}, observations have revealed that the PO/PN ratio is always $\geq$1. Our models of Section \ref{models} have however shown that the PO/PN ratio may become $\leq$1 for certain physical conditions and energetic phenomena. 

\begin{figure}
\begin{center}
\includegraphics[angle=0,width=0.45\textwidth]{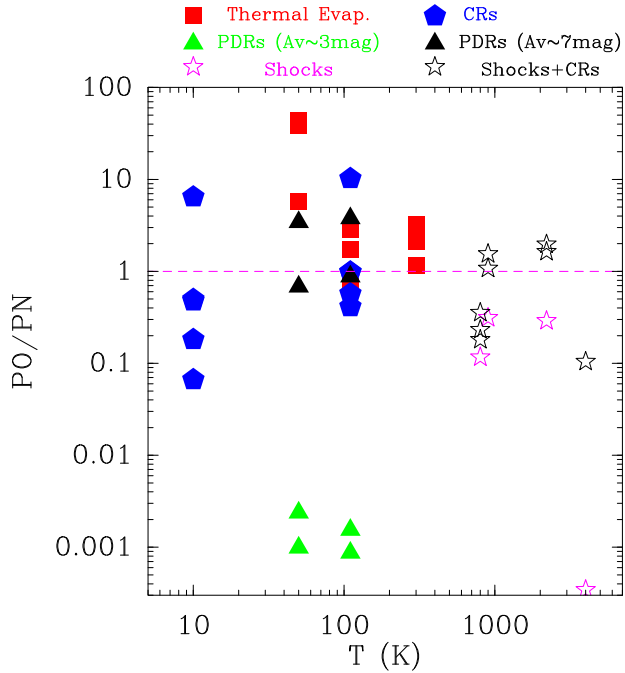}
\caption{Compilation of the PO/PN abundance ratios obtained within our models as a function of temperature. The ratios corresponding to two different time-scales are shown for every model: 10$^5$ yrs for models considering protostellar heating (red squares), cosmic rays (blue pentagons) and UV radiation (green and black triangles); and 10$^3$ yrs for models with shocks and shocks+cosmic rays (magenta and black empty stars). Horizontal dashed magenta line denotes the PO/PN ratio of 1.}
    \label{ratio-long}
\end{center}
\end{figure}

Figure \ref{ratio-long} presents the PO/PN ratios obtained by our models for a time-scale of 10$^5$ yrs \citep[typical of hot molecular cores and Galactic Center GMCs;][]{requena-torres06,rivilla16} for the models of proto-stellar heating, cosmic rays and UV radiation, and of 10$^3$ yrs \citep[typical of outflows; see e.g.][]{podio14} for the models with shocks and shocks with cosmic rays. All models consider a long-lived collapse phase.
As seen from Figure \ref{ratio-long}, the different energetic processes tend to cluster the PO/PN ratios toward different parts of the diagram. While proto-stellar heating and cosmic rays typically give PO/PN$\geq$1, this ratio is expected to be $<<$1 in PDRs (with A$_{\rm v}$$\sim$3 mag). This explains why the observed PO/PN ratios in hot molecular cores are $\geq$1 \citep{rivilla16}. For PDR regions with higher extinction (with A$_v$$\sim$7 mag), the chemistry of P-bearing species behaves in a similar way to that affected by photo-stellar heating, since the efficiency of molecular photo-dissociation drastically decreases at extinctions A$_v$$\geq$5 mag. 

For models with shocks, the PO/PN ratio is $\leq$1 in the absence of cosmic rays while this ratio is boosted above 1 under their effects. However, note that this trend may be a result of the incompleteness of the PO chemical network (see Section \ref{missing}). If the PO formation reaction proposed in Table \ref{tab-ratesPO} is experimentally or theoretically confirmed, the PO/PN ratios in shocks are also expected to be $\geq$1.      

Finally, for models with a short-lived collapse phase, the derived PO/PN ratios typically fall below 1, except for a few exceptions which involve enhancements of the cosmic-ray ionization rate by factors of 10$^2$ and 10$^4$.

\section{Conclusions}

P-bearing species such as PN and PO have started to be routinely detected in regions of the ISM affected by different energetic phenomena such as proto-stellar heating, cosmic rays, UV radiation and shock waves. In this paper, we have re-visited the chemistry of phosphorus in the ISM under energetic processing, putting special emphasis on the predicted ratio between PN and PO.

Our models show that a long-lived collapse is required to reproduce the PO/PN ratios $\geq$1 observed in hot molecular cores (see Section \ref{sec-ratio}). The models considering UV illumination confirm that P-bearing species are expected to be destroyed under the effects of strong UV radiation fields, as observed in the Orion Bar and Galactic Center GMCs (Section \ref{sec-UV}). Moderate UV radiation fields ($\chi$$\sim$100 Habing), however, should yield detectable abundances of P-bearing species such as PN, PO and PS. Models with enhanced Cosmic ray ionization rates provide a wide range of PO/PN ratios, although PN tends to be more abundant than PO. For models with C-type shock waves, an observed ratio PO/PN$\geq$1 is only obtained under the effects of enhanced rates of cosmic rays, unless the chemical network of PO is incomplete (Section \ref{sec-shock}). We propose that the reaction P + OH $\rightarrow$ PO + H (currently missing in all chemical databases) could be as efficient as its analogue reaction involving NO. Theoretical/experimental investigations are needed to establish the actual efficiency of this gas-phase formation route of PO.

\acknowledgments

We would like to thank the constructive comments from the referee, which helped to improve the original version of the manuscript. I.J.-S. and D.Q. acknowledge the financial support received from the STFC through an Ernest Rutherford Fellowship and Grant (proposals number ST/L004801 and ST/M004139). J.H. and S.V. acknowledge support from STFC (Grant number  ST/M001334/1).


\appendix

\section{Additional figures}
\begin{figure}
\begin{center}
\includegraphics[angle=0,width=0.6\textwidth]{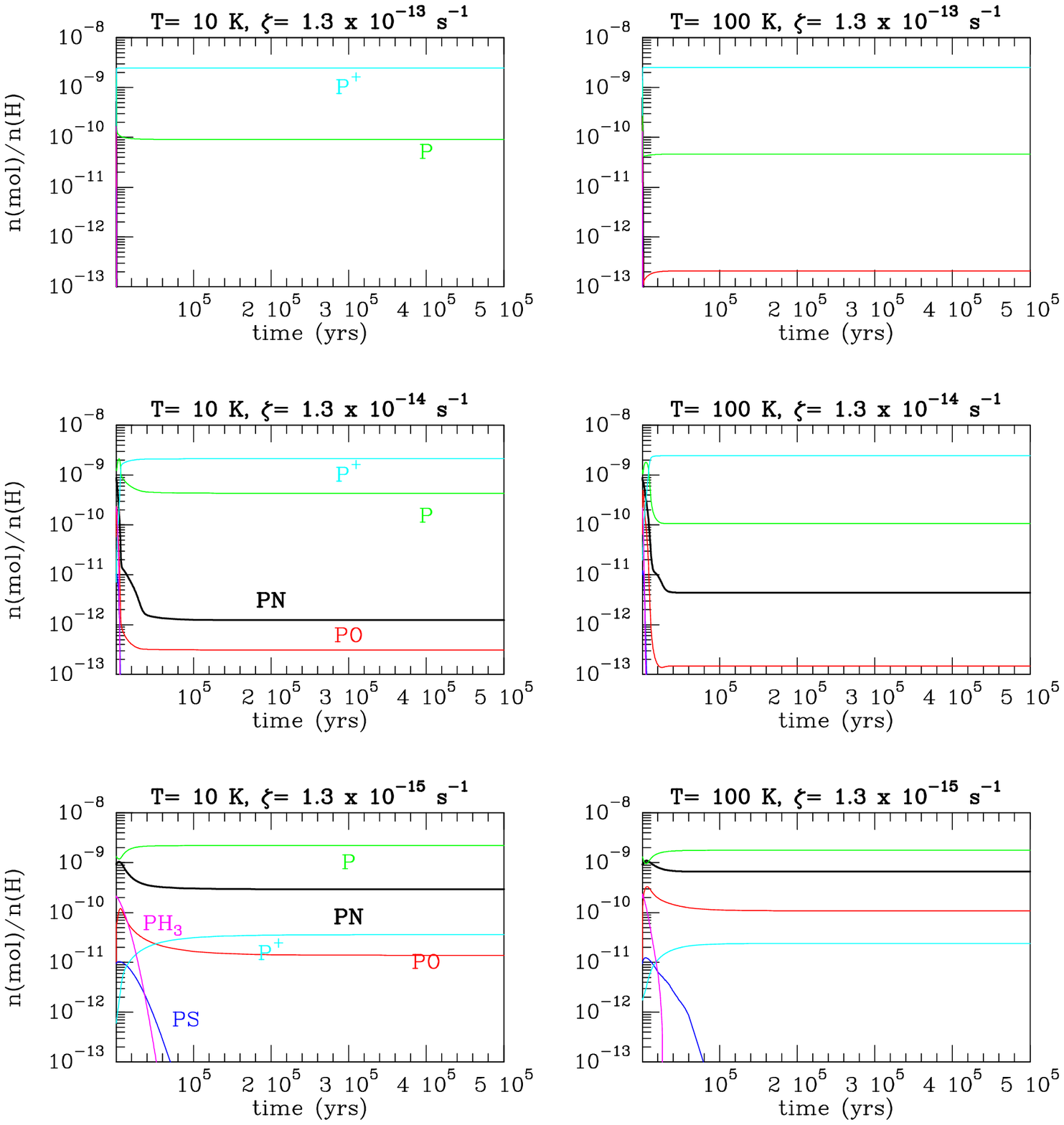}
\caption{Evolution of the abundances of P-bearing species as a function of time simulated for a hydrogen density n(H)=2$\times$10$^{4}$ cm$^{-3}$, enhanced cosmic-ray ionization rates $\zeta$ = 1.3$\times$10$^{-15}$ and 1.3$\times$10$^{-13}$ s$^{-1}$, and gas temperatures T = 10 K and 100 K, considering a short-lived phase for the collapse.}
    \label{CRs-1e4-short}
\end{center}
\end{figure}

\begin{figure}
\begin{center}
\includegraphics[angle=270,width=0.6\textwidth]{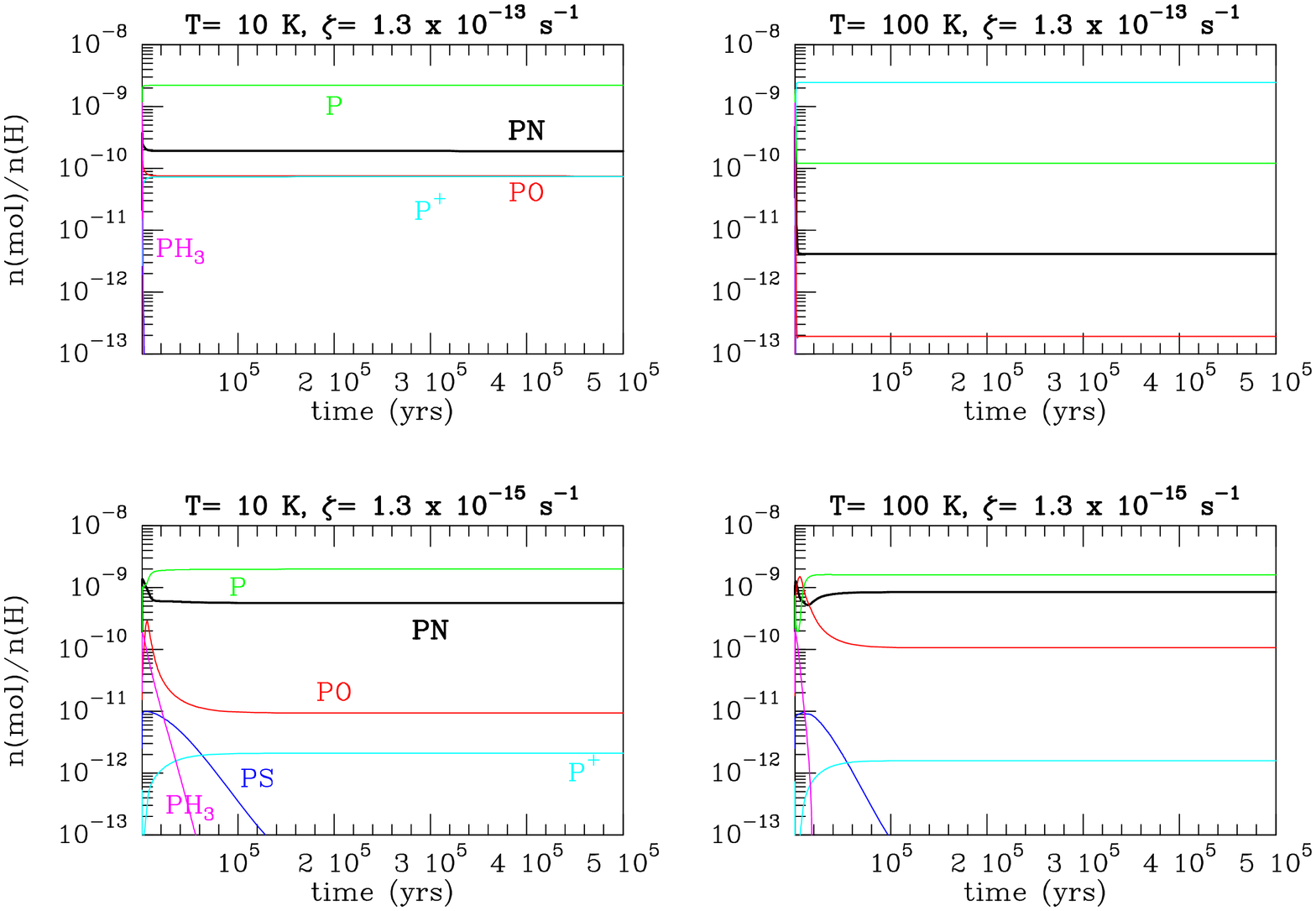}
\caption{Evolution of the abundances of P-bearing species as a function of time simulated for a hydrogen density n(H)=10$^{5}$ cm$^{-3}$, enhanced cosmic-ray ionization rates $\zeta$ = 1.3$\times$10$^{-15}$ and 1.3$\times$10$^{-13}$ s$^{-1}$, and gas temperatures T = 10 K and 100 K, considering a short-lived phase for the collapse.}
    \label{CRs-1e5-short}
\end{center}
\end{figure}

\begin{figure}
\begin{center}
\includegraphics[angle=270,width=1.0\textwidth]{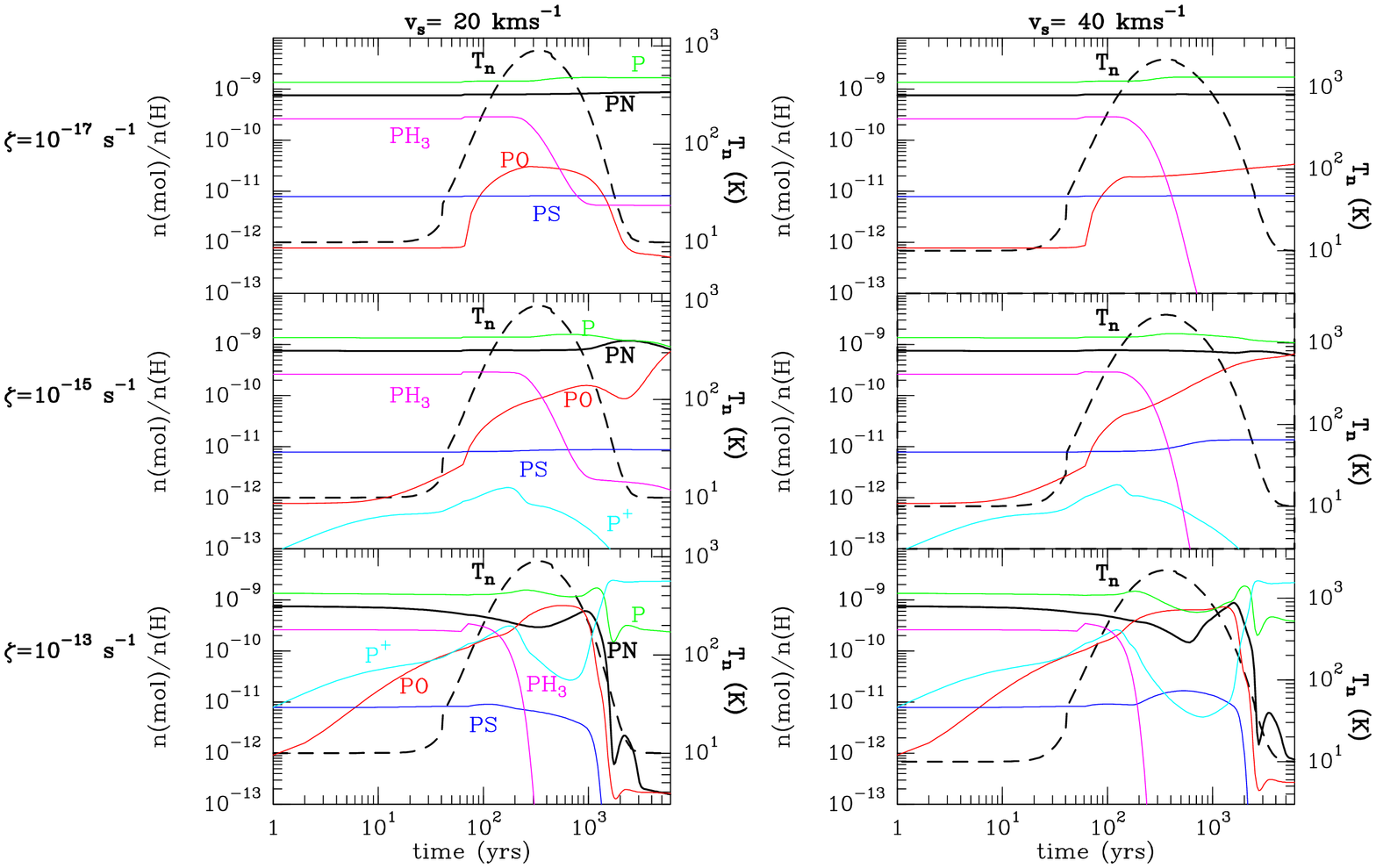}
\caption{Evolution of the abundances of P-bearing species as a function of time simulated for a hydrogen density n(H)=10$^{4}$ cm$^{-3}$, shock velocities v$_{s}$= 20 and 40 km s$^{-1}$, and cosmic-ray ionization rates $\zeta$ = 1.3$\times$10$^{-17}$, 1.3$\times$10$^{-15}$ and 1.3$\times$10$^{-13}$ s$^{-1}$. These models consider a short-lived phase for the collapse.}
    \label{shocks-1e4-short}
\end{center}
\end{figure}

\begin{figure}
\begin{center}
\includegraphics[angle=270,width=1.0\textwidth]{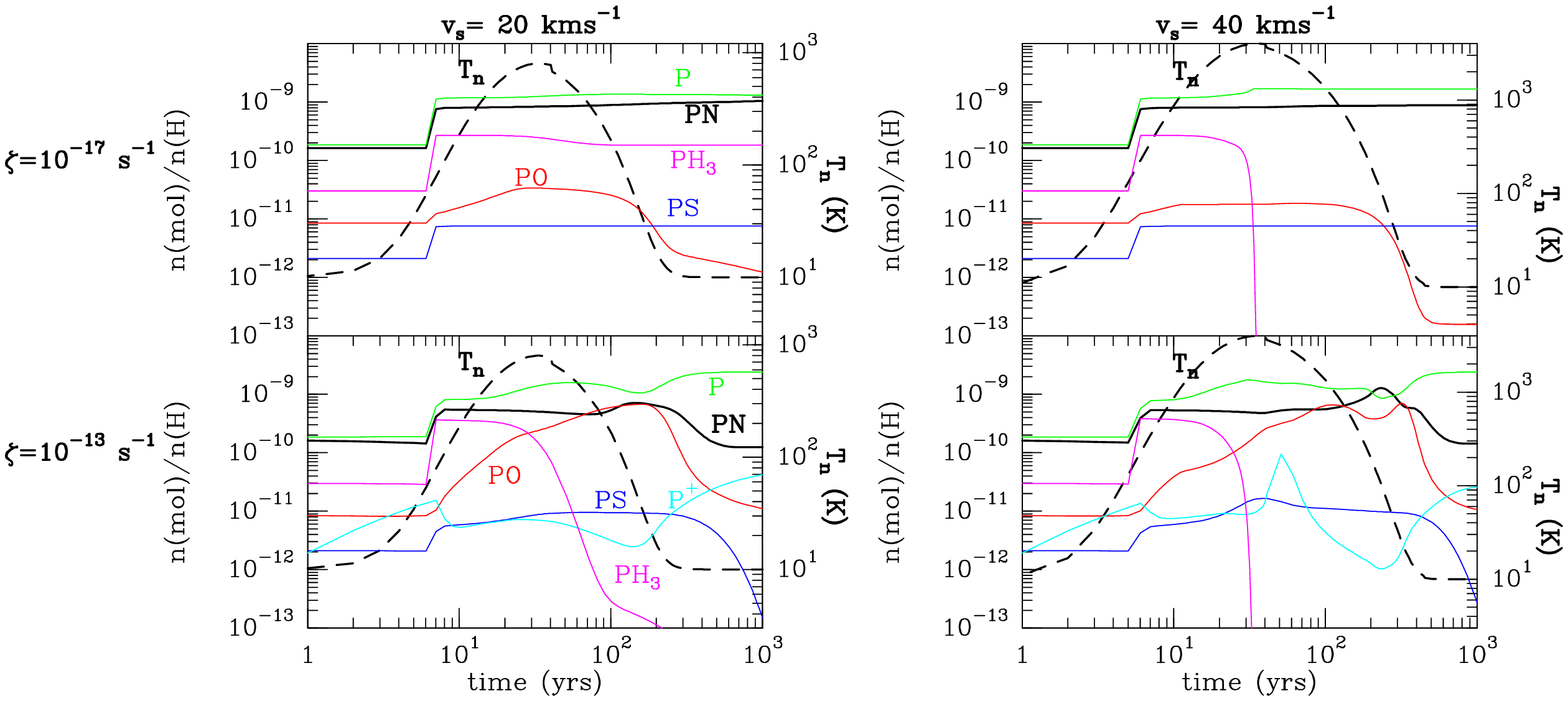}
\caption{Evolution of the abundances of P-bearing species as a function of time simulated for a hydrogen density n(H)=2$\times$10$^{5}$ cm$^{-3}$, shock velocities v$_{s}$= 20 and 40 km s$^{-1}$, and cosmic-ray ionization rates $\zeta$ = 1.3$\times$10$^{-17}$ and 1.3$\times$10$^{-13}$ s$^{-1}$. These models consider a short-lived phase for the collapse.}
    \label{shocks-1e5-short}
\end{center}
\end{figure}

In this Section, we present the figures for those models that consider a short-lived collapse phase and enhanced cosmic-ray ionization rates (Figures \ref{CRs-1e4-short} and \ref{CRs-1e5-short}), and shocks (Figures \ref{shocks-1e4-short} and \ref{shocks-1e5-short}).

\clearpage 

\begin{thebibliography}{}

\bibitem[\protect\citeauthoryear{Ag\'undez et al.}{2007}]{agundez07}
Ag\'undez, M., Cernicharo, J., \& Gu\'elin, M. 2007, ApJ, 662, L91A
\bibitem[\protect\citeauthoryear{Ag\'undez et al.}{2008}]{agundez08}
Ag\'undez, M., Cernicharo, J., Pardo, J. R., Gu\'elin, M., \& Phillips, T. G. 2008, A\&A, 485L, 33A
\bibitem[\protect\citeauthoryear{Ag\'undez et al.}{2014b}]{agundez14b}
Ag\'undez, M., Cernicharo, J., \& Gu\'elin, M. 2014b, A\&A, 570A, 45A
\bibitem[\protect\citeauthoryear{Ag\'undez et al.}{2014a}]{agundez14a}
Ag\'undez, M., Cernicharo, J., Decin, L., Encrenaz, P., \& Teyssier, D. 2014a, ApJ, 790, L27A
\bibitem[\protect\citeauthoryear{Anicich}{1993}]{anicich93}
Anicich, V. G. 1993, ApJ, 84, 215
\bibitem[\protect\citeauthoryear{Asplund et al.}{2009}]{asplund09}
Asplund, M., Grevesse, N., Sauval, A. J., \& Scott, P. 2009, ARA\&A, 47, 481A
\bibitem[\protect\citeauthoryear{Aota \& Aikawa}{2012}]{aota12}
Aota,T., \& Aikawa, Y., 2012, ApJ, 761, 74
\bibitem[\protect\citeauthoryear{Bohlin et al.}{1978}]{bohlin78}
Bohlin, R. C., Savage, B. D., \& Drake, J. F. 	
1978, ApJ, 224, 132
\bibitem[\protect\citeauthoryear{Caux et al.}{2011}]{caux11}
Caux, E., Kahane, C., Castets, A., et al.  2011, A\&A, 532A, 23C
\bibitem[\protect\citeauthoryear{Ceccarelli et al.}{2014}]{ceccarelli14}
Ceccarelli, C., Dominik, C., L\'opez-Sepulcre, A., Kama, M., Padovani, M., Caux, E., \& Caselli, P. 2014, ApJ, 790, L1
\bibitem[\protect\citeauthoryear{Cernicharo et al.}{2010}]{cernicharo10}
Cernicharo, J., et al. 2010, A\&A, 518, L136C
\bibitem[\protect\citeauthoryear{Charnley \& Millar}{1994}]{charnley94}
Charnley, S. B., \& Millar, T. J. 1994, MNRAS, 270, 570C
\bibitem[\protect\citeauthoryear{Codella et al.}{2012}]{codella12}
Codella, C., et al.  2012, ApJ, 759, L45C
\bibitem[\protect\citeauthoryear{Codella et al.}{2018}]{codella18}
Codella, C., et al. 2018, MNRAS, 474, 5694C
\bibitem[\protect\citeauthoryear{Cuadrado et al.}{2015}]{cuadrado15}
Cuadrado, S., Goicoechea, J. R., Pilleri, P., Cernicharo, J., Fuente, A., \& Joblin, C. 2015, A\&A, 575A, 82C
\bibitem[\protect\citeauthoryear{De Beck et al.}{2013}]{debeck13}
De Beck, E., Kaminski, T., Patel, N. A., Young, K. H., Gottlieb, C. A., Menten, K. M., \& Decin, L.  2013, A\&A, 558A, 132D
\bibitem[\protect\citeauthoryear{Fontani et al.}{2017}]{fontani17}
Fontani, F., et al. 2017, A\&A, 605A, 57F
\bibitem[\protect\citeauthoryear{Fontani et al.}{2016}]{fontani16}
Fontani, F., Rivilla, V.M., Caselli, P., et al., 2016, ApJ, 822, L30
\bibitem[\protect\citeauthoryear{Frerking et al.}{1982}]{frerking82}
Frerking, M. A., Langer, W. D., \& Wilson, R. W. 1982, ApJ, 262, 590
\bibitem[\protect\citeauthoryear{Goicoechea et al.}{2009}]{goicoechea09}
Goicoechea, J. R., Pety, J., Gerin, M., Hily-Blant, P., \& Le Bourlot, J.  2009, A\&A, 498, 771G
\bibitem[\protect\citeauthoryear{Goto et al.}{2014}]{goto14}
Goto, M., Geballe, T. R., Indriolo, N., Yusef-Zadeh, F., Usuda, T., Henning, T., \& Oka, T. 2014, ApJ, 786, 96G
\bibitem[\protect\citeauthoryear{Graedel et al.}{1982}]{graedel82}
Graedel, T. E., Langer, W. D., \& Frerking, M. A.  1982, ApJS, 48, 321G
\bibitem[\protect\citeauthoryear{Halfen et al.}{2008}]{halfen08}
Halfen, D. T., Clouthier, D. J., \& Ziurys, L. M. 2008, ApJ, 677, L101H
\bibitem[\protect\citeauthoryear{Harada et al.}{2010}]{harada10}
Harada, N., Herbst, E., \& Wakelam, V. 2010, ApJ, 721, 1570
\bibitem[\protect\citeauthoryear{Harada et al.}{2015}]{harada15}
Harada, N., Riquelme, D., Viti, S., et al. 2015, A\&A, 584A, 102H 
\bibitem[\protect\citeauthoryear{Holdship et al.}{2017}]{holdship17}
Holdship, J., Viti, S., Jim\'enez-Serra, I., Makrymallis, A., \& Priestley, F. 2017, AJ, 154, 38H
\bibitem[\protect\citeauthoryear{Jim\'enez-Serra et al.}{2008}]{jimenez08}
Jim\'enez-Serra, I., Caselli, P., Mart\'{\i}n-Pintado, J., \& Hartquist, T. W.  2008, A\&A, 482, 549J
\bibitem[\protect\citeauthoryear{Jim\'enez-Serra et al.}{2016}]{jimenez16}
Jim\'enez-Serra, I., et al. 2016, ApJ, 830, L6J
\bibitem[\protect\citeauthoryear{Jura et al.}{1978}]{jura78}
Jura, M., \& York, D. G. 1978, ApJ, 219, 861
\bibitem[\protect\citeauthoryear{Kaye \& Strobel}{1983}]{kaye83}
Kaye, J. A., \& Strobel, D. F. 1983, Geophys. Res. Letts., 10, 957
\bibitem[\protect\citeauthoryear{Koo et al.}{2013}]{koo13}
Koo, B.-C., Lee, Y.-H., Moon, D.-S., Yoon, S.-C., \& Raymond, J. C. 2013, Science, 342, 1346
\bibitem[\protect\citeauthoryear{Larson et al.}{1977}]{larson77}
Larson, H. P., Treffers, R. R., \& Fink, U. 1977, ApJ, 211, 972
\bibitem[\protect\citeauthoryear{Lee et al.}{1976}]{lee76}
Lee, J. H., Michael, J. V., Payne, W. A., Whylock, D. A., \& Stief, L. J. 1976, J. Chem. Phys., 65, 3210
\bibitem[\protect\citeauthoryear{Lefloch et al.}{2016}]{lefloch16}
Lefloch, B., et al.  2016, MNRAS, 462, 3937L
\bibitem[\protect\citeauthoryear{Lique et al.}{2018}]{lique18}
Lique, F., Jim\'enez-Serra, I., Viti, S., \& Marinakis, S. 2018, PCCP, 20, 5407L
\bibitem[\protect\citeauthoryear{Maci\'a et al.}{2005}]{macia05}
Maci\'a, E., 2005, Chem Soc. Rev., 34, 691
\bibitem[\protect\citeauthoryear{MacKay \& Charnley}{2001}]{mackay01}
MacKay, D. D. S.; Charnley, S. B.	 2001, MNRAS, 325, 545M
\bibitem[\protect\citeauthoryear{McElroy et al.}{2013}]{mcelroy13}	
McElroy, D., Walsh, C., Markwick, A. J., Cordiner, M. A., Smith, K., \& Millar, T. J.  2013, A\&A, 550A, 36M
\bibitem[\protect\citeauthoryear{Milam et al.}{2008}]{milam08}
Milam, S. N., Halfen, D. T., Tenenbaum, E. D., Apponi, A. J., Woolf, N. J., \& Ziurys, L. M. 2008, ApJ, 684, 618M
\bibitem[\protect\citeauthoryear{Millar}{1991}]{millar91}
Millar, T. J. 1991, A\&A, 242, 241M
\bibitem[\protect\citeauthoryear{Mininni et al.}{2018}]{mininni18}
Mininni, C., Fontani, F., et al. 2018, MNRAS, accepted, arXiv:1802.00623
\bibitem[\protect\citeauthoryear{Molinari et al.}{2000}]{molinari00}
Molinari, S., Brand, J., Cesaroni, R., \& Palla F., 2000, A\&A, 355, 617
\bibitem[\protect\citeauthoryear{Morton et al.}{1974}]{morton74}
Morton, D. C. 1974, ApJ, 193, L35M 
\bibitem[\protect\citeauthoryear{Neufeld et al.}{2005}]{neufeld05}
Neufeld, D. A., Wolfire, M. G., \& Schilke, P. 2005, ApJ, 628, 260N
\bibitem[\protect\citeauthoryear{Pagani et al.}{2012}]{pagani12}
Pagani, L., et al. 2012, Philosophical Transactions of the Royal Society A, 370, 1978, 5200
\bibitem[\protect\citeauthoryear{Podio et al.}{2014}]{podio14}
Podio, L., Lefloch, B., Ceccarelli, C., Codella, C., \& Bachiller, R. 2014, A\&A, 565A, 64P
\bibitem[\protect\citeauthoryear{Qu\'enard et al.}{2018}]{quenard18}
Qu\'enard, D., Jim\'enez-Serra, I., Viti, S., Holdship, J., \& Coutens, A. 2018, MNRAS, 474, 2796 
\bibitem[\protect\citeauthoryear{Requena-Torres et al.}{2006}]{requena-torres06}
Requena-Torres, M. A., Mart\'{\i}n-Pintado, J., Rodr\'{\i}guez-Franco, A., Mart\'{\i}n, S., Rodr\'{\i}guez-Fern\'andez, N. J., \& de Vicente, P. 2006, A\&A, 455, 971R
\bibitem[\protect\citeauthoryear{Rivilla et al.}{2016}]{rivilla16}
Rivilla, V. M., Fontani, F., Beltr\'an, M. T., Vasyunin, A., Caselli, P., Mart\'{\i}n-Pintado, J., \& Cesaroni, R.  2016, ApJ, 826, 161R
\bibitem[\protect\citeauthoryear{Rivilla et al.}{2018}]{rivilla18}
Rivilla, V. M., Jim\'enez-Serra, I., Zeng, S., et al. 2018, MNRAS, tempL2R 
\bibitem[\protect\citeauthoryear{Roberts et al.}{2007}]{roberts07}	
Roberts, J. F., Rawlings, J. M. C., Viti, S., \& Williams, D. A. 2007, MNRAS, 382, 733R
\bibitem[\protect\citeauthoryear{Schilke et al.}{1995}]{schilke95}
Schilke, Peter, Phillips, T. G., \& Wang, Ning  1995, ApJ, 441, 334S
\bibitem[\protect\citeauthoryear{Schilke et al.}{2001}]{schilke01}
Schilke, P., Pineau des For\^ets, G., Walmsley, C. M., \& Mart\'{\i}n-Pintado, J. 2001, A\&A, 372, 291S
\bibitem[\protect\citeauthoryear{Smith et al.}{1989}]{smith89}
Smith, D., McIntosh, B. J., \& Adams, N. G. 1989, J. Chem. Phys. 90, 6213 
\bibitem[\protect\citeauthoryear{Tenenbaum et al.}{2007}]{tenenbaum07}
Tenenbaum, E. D., Woolf, N. J., \& Ziurys, L. M. 2007, ApJ, 666, L29T
\bibitem[\protect\citeauthoryear{Tenenbaum \& Ziurys}{2008}]{tenenbaum08}
Tenenbaum, E. D., \& Ziurys, L. M. 2008, ApJ, 680, L121T
\bibitem[\protect\citeauthoryear{Thorne et al.}{1983}]{thorne83}
Thorne, L. R., Anicich, V. G., \& Huntress, W. T. 1983, Chem. Phys. Lett., 98, 162
\bibitem[\protect\citeauthoryear{Thorne et al.}{1984}]{thorne84}
Thorne, L. R., Anicich, V. G., Prasad, S. S., \& Huntress, W. T., Jr. 1984, ApJ, 280, 139
\bibitem[\protect\citeauthoryear{Turner et al.}{1990}]{turner90}
Turner, B. E., Tsuji, T., Bally, J., Guelin, M., \& Cernicharo, J. 1990, ApJ, 365, 569T
\bibitem[\protect\citeauthoryear{Turner et al.}{1999}]{turner99}
Turner, B. E., Terzieva, R., \& Herbst, Eric 1999, ApJ, 518, 699
\bibitem[\protect\citeauthoryear{Viti \& Williams}{1999}]{viti99}
Viti, S., \& Williams, D. A. 1999, MNRAS, 310, 517V
\bibitem[\protect\citeauthoryear{Viti et al.}{2004}]{viti04}	
Viti, S., Collings, M. P., Dever, J. W., McCoustra, M. R. S., \& Williams, David A. 2004, MNRAS, 354, 1141V
\bibitem[\protect\citeauthoryear{Wakelam et al.}{2017}]{wakelam17}
Wakelam, V., Loison, J.-C., Mereau, R., \& Ruaud, M. 2017, MolAs, 6, 22W
\bibitem[\protect\citeauthoryear{Weisstein \& Serabyn}{1994}]{weisstein94}	
Weisstein, E. W., \& Serabyn, E. 1994, Icarus, 109, 367W
\bibitem[\protect\citeauthoryear{Yamaguchi et al.}{2011}]{yamaguchi11}
Yamaguchi, T., Takano, S., Sakai, N., et al., 2011, PASJ, 

\end{thebibliography}
\end{document}